\documentclass[journal,twoside,web]{ieeecolor}

\usepackage{generic}
\usepackage{amsmath,amssymb,amsfonts}
\usepackage{algorithmic}
\usepackage{graphicx}
\usepackage{algorithm,algorithmic}
\usepackage{textcomp}
\usepackage[comma,numbers,square,sort&compress]{natbib}

\usepackage{subfigure}
\usepackage{epstopdf}

\usepackage{amssymb}
\usepackage{amsmath}
\usepackage{algorithm}
\usepackage{etoolbox}
\usepackage{mathrsfs}

\usepackage{booktabs}
\usepackage{multirow}
\usepackage{array}
\usepackage{makecell}
\usepackage{hyperref} 
\hypersetup{
	colorlinks=true, % 启用彩色链接
	linkcolor=blue,  % 内部链接颜色
	citecolor=blue,  % 引用链接颜色
	filecolor=blue,  % 文件链接颜色
	urlcolor=blue    % URL 链接颜色
}
\let\classAND\AND
\let\AND\relax
\let\AND\classAND
	\AtBeginEnvironment{algorithmic}{\let\AND\algoAND}
	
	\floatname{algorithm}{Algorithm}

	\newtheorem{remark}{Remark}
	\newtheorem{theorem}{Theorem}
	\newtheorem{lemma}{Lemma}
	\newtheorem{problem}{Problem}
	\newtheorem{assumption}{Assumption}
	\newtheorem{definition}{Definition}
	\usepackage{xcolor} % 引入颜色处理包
	\definecolor{IEEEblue}{rgb}{0,0.2,0.6} 
	
	\def\BibTeX{{\rm B\kern-.05em{\sc i\kern-.025em b}\kern-.08em
			T\kern-.1667em\lower.7ex\hbox{E}\kern-.125emX}}
	\makeatletter
	
\begin{document}
		\title{Byzantine-Resilient Output Optimization of Multiagent via Self-Triggered Hybrid Detection Approach} 
		\author{Chenhang Yan, \IEEEmembership{Graduate Student Member, IEEE}, Liping Yan, Yuezu Lv, \IEEEmembership{Senior Member, IEEE},\\ Bolei Dong, and Yuanqing Xia, \IEEEmembership{Fellow, IEEE}
		\thanks{This work was supported by the Beijing Natural Science Foundation under Grant 4242049, and the National Natural Science Foundation of China under Grant 62073036, Grant 62088101, Grant 62273045, Grant U2341213, and the Beijing Nova Program under Grant 20230484481. \emph{(Corresponding author: Liping Yan.)}}
			\thanks{C. Yan, L. Yan, B. Dong and Y. Xia are with School of Automation, Beijing Institute of Technology, Beijing 100081, P.~R.~China. (e-mail: chenhangyan\_mail@163.com; ylp@bit.edu.cn; bldong168@163.com; xia\_yuanqing@bit.edu.cn).}
			\thanks{Y. Lv is with MIIT Key Laboratory of Complex-field Intelligent Sensing, Beijing Institute of Technology, Beijing, 100081, China. (e-mail: yzlv@bit.edu.cn)}
		}
		\maketitle
		
		\begin{abstract} How to achieve precise distributed optimization despite unknown attacks, especially the Byzantine attacks, is one of the critical challenges for multiagent systems.
			This paper addresses a distributed resilient optimization for linear heterogeneous multi-agent systems faced with adversarial threats. We establish a framework aimed at realizing resilient optimization for continuous-time systems by incorporating a novel self-triggered hybrid detection approach. The proposed hybrid detection approach is able to identify attacks on neighbors using both error thresholds and triggering intervals, thereby optimizing the balance between effective attack detection and the reduction of excessive communication triggers. Through using an edge-based adaptive self-triggered approach, each agent can receive its neighbors' information and determine whether these information is valid. If any neighbor prove invalid, each normal agent will isolate that neighbor by disconnecting communication along that specific edge. Importantly, our adaptive algorithm guarantees the accuracy of the optimization solution even when an agent is isolated by its neighbors.
		\end{abstract}
		
		%\allowdisplaybreaks[2]
		\section{Introduction}
		\label{sec:introduction}
		\IEEEPARstart{D}{istributed} consensus control of multiagent systems (MASs) has emerged as a prominent field of research over the past two decades. Specifically, the optimal consensus control, commonly referred to as distributed optimization, seeks to minimize a cost function for a MAS through the utilization of local information \cite{nedic2010constrained}. This method demonstrates significant potential and practical relevance, enhancing MAS capabilities in navigating complex environments.
		
	 With the advancement of cyber-physical systems, security challenges such as Byzantine attacks and denial-of-service attacks have become increasingly critical. Robust security measures are essential to protect the integrity and functionality of MAS operations. Due to their open communication and distributed nature, MASs are particularly vulnerable to cyber attacks, which will compromise the stability of the systems. Adversarial agents, containing malicious or Byzantine agents, may breach established control protocols and transmit misleading information to their neighbors, thereby disrupting the objectives of the whole network. So far, the importance of distributed resilient control for MASs has been widely recognized and has gained much attention by researchers \cite{ishii2022overview,ramos2023discrete,luo2024resilient}.
	 	
	 	\subsection{Related Works and Motivations}
	 	To tackle the resilient control for MASs under attacks, there are generally two mainstream methods. The first method is to establish a protocol based on the potential tampered information from neighbors to mitigate the influence caused by the adversarial agents \cite{gusrialdi2018competitive,mustafa2020resilient, zegers2021event,sadabadi2023resilient}. However, this method primarily seeks to estimate and offset the attacked signal in the system, minimizing the impact of attacks and achieving consensus with an error bound. Therefore, it is particularly effective in leader-following consensus scenarios, ensuring followers converge within a bounded region around the leader. Conversely, the second method is designing a mechanism to employ the receive data in an acceptability region, which can be roughly clarified into two categories: the mean-subsequence-reduced (MSR)-based method, and the detection-based method. The MSR-based method relies on a ($2r+1$)-robustness communication graph, under which each agent can reach consensus in the presence of at most $r$ total/local adversarial attackers. Also, the final states of the normal agents will converge into the convex hull of their initial states. The resilience of MASs under MSR-based consensus control has been detailed in \cite{leblanc2013resilient, dibaji2017resilient, wen2023joint, usevitch2019resilient,yan2024dynamic,yan2022resilient}, while resilient optimization of MASs has been covered in \cite{su2020byzantine,wu2023byzantine,kuwaranancharoen2020byzantine}. With regard to resilient optimization, however, the optimal solution of MASs under such method will converge to an approximation constrained by the aggregate objective function of all normal agents. As for the detection-based method, it first identifies whether the agent is under attack and then isolates or weaken the attacked one from the normal one \cite{barboni2020detection, zhao2021resilient,hu2023detection}.  These strategies will cut off the communication to the abnormal agents once identified, thereby directly removing the impact of attacked agents on the system. 
	 	
	 	Up to now, the distributed resilient optimization of MASs under adversarial agents have made some progress \cite{su2020byzantine,wu2023byzantine, kuwaranancharoen2020byzantine}, \cite{fu2021resilient}, \cite{zhao2019resilient}. Nevertheless, the common features of them are evident in that the system dynamics of them are first-order and discrete-time, and these approaches can only lead the agents to an approximate sub-optimal solution that includes the global minimizer. Notably, the MSR-based resilient optimization approaches presented in \cite{su2020byzantine,wu2023byzantine, kuwaranancharoen2020byzantine}, \cite{fu2021resilient} have strict requirements on communication edges, while \cite{zhao2019resilient} designs a resilient optimization protocol relying on a specific trusted agent in networks. Consequently, how to ensure all the normal agents to estimate their optimal solutions without affected by the adversarial agents becomes critical. Furthermore, in practical MASs, the dynamics are often high-order and nonidentical. Therefore, developing a comprehensive framework for general heterogeneous MASs (HMASs) to address resilient optimization challenges warrants thorough investigation. On the other hand, event-triggered mechanism for MASs has been achieved great attention recently \cite{cheng2023discrete}, \cite{garcia2016periodic}, not only due to it can reduce communication frequency, but also have the potential uses in abnormal detection of agents. The result in \cite{zegers2021event} demonstrates that event-triggered detection-based consensus protocol is able to distinguish if there are misbehaving agents among neighbors, where a threshold is carried out for attack detection. However, such threshold for the agent in \cite{zegers2021event} still requires untrustworthy parameters from neighbors, which thereby greatly increases the risk of attacks on these parameters. Moreover, it also requires the eigenvalues of the Laplacian matrix, which, as a global knowledge, cannot be readily obtainable on some occasions.
	 	
	  \subsection{Contributions and Organization}
	 	Motivated by the aforementioned challenges, the paper proposes a framework of an adaptive event-triggered hybrid detection based distributed resilient optimization protocol for HMASs in scenarios involving adversarial agents. This framework ensures that all normal agents are capable of converging to the global optimal value under adversarial conditions.
	 	
	 	The main innovations of the article are as follows:
	 	\begin{itemize}
	 		\item A hybrid detection-based event-triggered protocol is proposed to achieve distributed optimization, ensuring that the final optimal solution is consistent across all normal agents. The designed detection method incorporates two types of time-varying detection thresholds, i.e., trigger interval threshold and trigger error threshold, to increase the success rate of attack detection.
	 		\item We have developed an innovative adaptive optimization protocol that operates effectively independent of the non-zero smallest eigenvalue of the Laplacian matrix. This feature is crucial, as the eigenvalue information, which is influenced by the isolation of adversarial agents, is unknown to other agents.
	 		\item Utilizing the designed event-triggered protocol, the communication between agents is discrete, and the existence of the positive minimal event-triggered interval (MEI) is guaranteed. In addition, this protocol is the first to explore a hybrid event-triggered mechanism for distributed optimization.
	 	\end{itemize}
 	
	 	The rest of this article is organized as follows. In section II, the knowledge of graph theory, lemmas, the attack model, and the description of the problem are given. In sections III and IV, the designed resilient protocol and the attack detection and isolation algorithm are illustrated, respectively. The stability analysis of this paper is stated in Section V, and a simulation example is carried out in Section VI. Finally, Section VII offers a conclusion,  Section VIII includes appendices.

		\textbf{Notations:} $\mathbb{R}^n$ and $\mathbb{R}^{n \times m}$ signifies the Euclidean spaces of $n\times 1$ the $n\times m$ real matrices, respectively. $\mathbb{Z}^+$ represents the set of positive integers. For ${X} \in \mathbb{R}^{n \times n}$, denote $\lambda_m({X})$, $\lambda_M({X})$ as the smallest and the largest positive real parts of all eigenvalues of $X$, respectively.  $\mathbb{Z}^+$ denotes the positive number set. The notation $\mathrm{col}\left\{ a_1,a_2,...,a_N \right\} =\left[ a_{1}^{T},a_{2}^{T},...,a_{N}^{T} \right] ^T$ stands for a column vector formed by stacking all vectors $a_i \in \mathbb{R}^{m}$,  $i=1,..,N$. The function $\rm{diag}\{\cdot\}$ represents a block matrix composed of all the arguments, and  the symbol $\|\cdot\|$ means the $2$-norm of a matrix.
		
		\section{Preliminaries}
		\subsection{Graph Theory and Supporting Lemmas}
		An unsigned weighted undirected graph among $N$ nodes is described as $\mathcal{G} =\left\{ \mathcal{V}, \mathcal{E} \right\}$. $\mathcal{V} =\left\{ v_1,...,v_N \right\}$ represents the set of $N$ nodes, and $\mathcal{E} \subset \mathcal{V} \times \mathcal{V}$ means the set of the undirected edges connected to nodes,
		where $\mathcal{E}(i, j)\in \mathcal{E}$ denotes the edge connecting agents $i$ and $j$.
		
		We assume that in the initial time ($i.e., t=0$), all agents in $\mathcal{V}$ behave normally. Then define $\mathcal{N}$ as the number of the agents which is not isolated by any other agent at the current time, while ${\mathcal{Q}}$ represents the number of the misbehaving agents that have been isolated so far. In addition, the abovementioned parameters satisfies condition $N=\mathcal{N}+{\mathcal{Q}}$, where $\mathcal{N}=N, {\mathcal{Q}}=0$ when $t=0$. Once any agent is isolated from the group of normal agents, the parameters $\mathcal{N}$ and ${\mathcal{Q}}$ will be updated immediately. Subsequently, let $\mathcal{G}_\mathcal{N}$ denote the graph composed of the agents that are not isolated by any other normal one. For later expression, we denote $\mathcal{I}_\mathcal{N}= \{1,..,\mathcal{N}\}$ and $\mathcal{I}_\mathcal{B}= \{\mathcal{N}+1,..,N\}$ as the index of all normal agents and all misbehaving agents, respectively.
		
	Let $\mathcal{A} =\left[ a_{ij} \right] \in \mathbb{R} ^{\mathcal{N}\times \mathcal{N}}$ stand for the weighted adjacent matrix of graph $\mathcal{G}$ constructed by $\mathcal{N}$ normal agents, in which $a_{ij} =a_{ji} \geqslant 0$ is the weight of each communication edge $\mathcal{E}(i, j)$. %If node $j$ can broadcast information to node $i$, we have that $a_{ij}>0$, node $i$ is the out-neighbor of node $j$ and correspondingly, node $j$ is the in-neighbor of node $i$; otherwise, $a_{ij}=0$, and those two nodes are not the in-neighbor or out-neighbor to each other. Especially, a graph is characterized as undirected if the condition $a_{ij}=a_{ji}$ always holds.
	If node $j$ can transmit message to node $i$, the corresponding directed path is denoted as $\mathcal{P} \left( i,j \right)$.
	Let $\mathcal{N} _{i}$ be the set of all neighbors of node $i$.  A undirected graph is called connected if for any pair of nodes $j_1$ and $j_2$ in the graph, there is a path between $j_1$ and $j_2$; otherwise, it is disconnected. The concept of in-degree of node $i$ is described by $d_{i}=\sum_{j\in \mathcal{N} _{i}}{a_{ij}}$. Then the Laplacian matrix $L=\left[ L_{ij} \right] \in \mathbb{R} ^{\mathcal{N} \times \mathcal{N}}$ is given by $L=D-\mathcal{A}$ where $D=\mathrm{diag}\left\{ d_{1},...,d_{\mathcal{N}} \right\}$.
		\begin{lemma} \textit{\cite{li2013distributed}} \label{lemma.0} 
			For an undirected graph $\mathcal{G}$, if it is connected, then all eigenvalues of $L$ have nonnegative real parts, and $0$ is the only zero eigenvalue,  corresponding to the right eigenvector $\mathbf{1}$. The smallest positive eigenvalue $\lambda _2$  of $L$ is given by $\lambda _2=\min_{\nu \ne 0,\mathbf{1}^T\nu =0} \frac{\nu  ^TL\nu}{\nu^T\nu}$.
		\end{lemma}

\begin{lemma}(\textit{Filippov Solution \cite{clarke1990optimization}})\label{lemma.5} 
	Let $\dot{z}=\hat{f}\left( z,t \right)$ be a differential function, where $z\in \mathbb{R} ^n$ is absolutely continuous in $\left[ 0,T_1 \right)$, $T_1>0$, and $\hat{f}$ is measurable, essential locally bounded.
	Let $z$ be a Filippov solution of the differential inclusion $\dot{z}\in \mathbb{F} \left( z \right)$, where $\mathbb{F} \left( z \right):\mathbb{R} ^n\rightarrow \mathbb{R} ^n$ is the set-valued map, denoted as $\mathbb{F} \left( z \right) =\bigcap\nolimits_{\varpi >0}^{}{\bigcap\nolimits_{\varsigma \left( \mathscr{X} \right) =0}^{}\overline{cov}}\hat f\left( \mathcal{B} _o\left( z, \hat \varpi \right) \setminus \mathscr{X} \right)$, where $\varsigma (\mathscr{X})$ signifies the Lebesgue measure of the set $\mathscr{X}$; $\bigcap\nolimits_{\varsigma( \mathscr{X} ) =0}$ stands for the intersection of all sets $\mathscr{X}$ whose measures are zero; $\mathcal{B} _o\left( z,\hat \varpi \right)$ means the open ball centered at $z$ with $\hat \varpi$ as the radius; the function $\overline{cov}\left( \cdot \right)$ represents the convex closure.
\end{lemma}

Let $V: \mathbb{R} ^n\rightarrow \mathbb{R}$ be locally Lipschitz and continuous, and $\partial V\left( z \right) =\overline{cov} \left\{ \lim _{l\rightarrow \infty}\nabla V\left( z_l \right) |z_l\rightarrow z,z_l\in \varDelta \left( V \right) \cup \mathscr{X} \right\}$ be Clarke's generalized gradient with $\varDelta \left( V \right) \subset \mathbb{R} ^n$ being the point set where $V$ is not differentiable. Based on this, define the set-valued Lie derivative of $V$ relative to $\hat f$ as $\ell \dot{V}=\left\{ \bar{\mu}\in \mathbb{R} |\bar{\mu}=\bar{\varrho}^T\bar{\varepsilon},\bar{\varrho}\in \mathbb{F} \left( z \right), \bar{\varepsilon}\in \partial V \right\}$ \cite{filippov2013differential}.
\begin{lemma}\cite{clarke1990optimization}\label{lemma.6}
	Given a set $Q$ satisfies that $Q=\left\{ z\in \mathbb{R} ^n|\left\| z \right\| \leqslant r, \forall r>0 \right\}$. If a regular function $V(z): \mathbb{R}^n \rightarrow \mathbb{R}$ satisfies that for $z=0$, $V(z)=0$, and for $z\ne 0$, $0<V_a\left( \left\| z \right\| \right) \leqslant V\left( z \right) \leqslant V_b\left( \left\| z \right\| \right)$, where $z \in Q$ and $V_a\left( \cdot \right), V_b\left( \cdot \right)$ belong to class $\mathcal{K}$, then $\ell \dot{V}\left( z \right) \leqslant 0,z\in Q$ indicates that $z$ converges to zero as $t \rightarrow \infty$.
\end{lemma}

\subsection{Attack Model}
This article aims to address the case where the node set in a network includes some abnormal agents. For this reason, we need the following essential definitions.
\begin{definition}
	In graph $\mathcal{G}$, an agent is named a normal agent if it transmits its real information to all other neighbors simultaneously; an agent is called a Byzantine agent if it transmits different tampered information to different neighbors at different times.
\end{definition}
\begin{definition}\textit{\cite{leblanc2013resilient}($r$-local/total Byzantine)}
	If each agent among graph $\mathcal{G}$ has at most $r$ Byzantine neighbors, then $\mathcal{G}$ is called $r$-local, i.e., $\left| \mathcal{N}_{i}\cap \mathcal{B} \right|\leqslant r$, in which $\mathcal{B}$ stands for the set of Byzantine agents in $\mathcal{G}$. If graph $\mathcal{G}$ consists of at most $r$ Byzantine agents, then $\mathcal{G}$ is referred to as $f$-total, i.e., $\left| \mathcal{B} \right|\leqslant r$.
\end{definition}
\begin{definition}\textit{\cite{leblanc2013resilient}($r$-reachable set)}
	For any nonempty agent set $\mathcal{X} \subset \mathcal{V}$, $\mathcal{X}$ is called an $r$-reachable set if there exists at least one agent in set $\mathcal{X}$ which has $r$ neighbors outside of $\mathcal{X}$, i.e., $\left| \mathcal{N}_{i}\setminus \mathcal{X} \right|\geqslant r$.
\end{definition}

Due to the existence of the Byzantine agents in networks, the normal agent may receive falsified data from neighboring agents. To describe this situation, we introduce the symbol $\theta_{ij}\in \mathbb{R}^{n_i}$ to represent any received variable transmitted from agent $j$ to agent $i$.
For agent $i$, since its neighbor $j$ may be the Byzantine agent, we can model $\theta_{ij}$ as 
\begin{align}
	\theta_{ij}(t)=&\theta_j(t)+\epsilon _{ij}^{\theta}(t),\, j \in \mathcal{N}_i
\end{align}
where $\theta_j\in \mathbb{R}^{n_i}$ denotes the true variable of agent $j$, which is not affected by attackers; $\epsilon_{ij}^{\theta}\in \mathbb{R}^{n_i}$ represents the unknown deviation of $\theta_j$ on path $\mathcal{P}(i,j)$, which is tampered with by attackers. Conversely, if agent $j$ is a normal agent, then it follows that $\theta_{ij}=\theta_j$ and $\epsilon _{ij}^{\theta}=0$. 

\subsection{Problem Formulation}
		To begin with, consider a team of $N$ agents with nonidentical system dynamics modeled by
		%\begin{equation}
		\begin{align}\label{e1}
			\dot{x}_i(t)=&A_ix_i(t)+B_iu_i(t)\nonumber
			\\
			y_i(t)=&C_ix_i(t)
		\end{align}
		where $i \in \mathcal{I}$, $\mathcal{I}=\{1, ..., N\}$ denotes the index of agents; $x_i(t) \in \mathbb{R}^{n_i}$, $y_i(t) \in \mathbb{R}^{q}$, $u_i(t) \in \mathbb{R}^{p_i}$ are, respectively, the state, the output and the control input of agent $i$; the matrices $A_i, B_i, C_i$ are of appropriate dimensions, and $C_i$ is of full column rank. 
		
We present the structure of distributed controller as
\begin{align}\label{e1a}
u_i=&\psi _i\left( x_i,\delta _i,w_i,\delta _{ij},w_{ij} \right) 
\nonumber\\
\dot{\delta}_i=&\psi _{i}^{\delta}\left( \delta _i,w_i,\delta _{ij},w_{ij} \right) 
\nonumber\\
\dot{w}_i=&\psi _{i}^{w}\left( \delta _i,\delta _{ij} \right),\ \ \forall j\in \mathcal{N} _i
\end{align}
where  $i \in \mathcal{I}$, the operators $\psi_i$, $\psi_i^{\delta}$, $\psi_i^{w}$ represent nonlinear functions to be designed, and $w_i \in \mathbb{R}^{n_i}$, $\delta_i \in \mathbb{R}^{n_i}$ serve as the internal variables of agent $i$, while $\delta_{ij} \in \mathbb{R}^{n_i}$, $w_{ij} \in \mathbb{R}^{n_i}$ are the external variables transmitted from agent $j$ to agent $i$ with $\delta _{ij}=\delta _j+\epsilon _{ij}^{\delta},w_{ij}=w_j+\epsilon _{ij}^{w}$. All  the above variables will be determined later.		
		
		For the HMAS (\ref{e1}), there always exists a matrix $T \in \mathbb{R}^{n_i \times p_i}$ such that $TB_i= \mathrm{col}\{0,  \bar B_{i}\}$ with $\bar B_{i}$ being a full rank matrix, i.e., $\mathrm{rank}( \bar{B}_i) =p_i$. Without loss of generality, the input matrix $B_i$ in (\ref{e1}) is assumed to be $\mathrm{col}\{0, \bar B_{i}\}$ in subsequent analysis. From this basis, system (\ref{e1}) can be partitioned as
		\begin{align}\label{e2b}
			\left[ \begin{matrix}
				\dot{x}_{i,1}\\
				\dot{x}_{i,2}\\
			\end{matrix} \right] =&\left[ \begin{matrix}
				A_{i,11}&		A_{i,12}\\
				A_{i,21}&		A_{i,22}\\
			\end{matrix} \right] \left[ \begin{matrix}
				x_{i,1}\\
				x_{i,2}\\
			\end{matrix} \right] +\left[ \begin{matrix}
				0\\
				\bar{B}_i\\
			\end{matrix} \right] u_i
			\nonumber\\
			y_i=&\left[ \begin{matrix}
				C_{i,1}&		C_{i,2}\\
			\end{matrix} \right] \left[ \begin{matrix}
				x_{i,1}\\
				x_{i,2}\\
			\end{matrix} \right] 
		\end{align}
		\begin{assumption}\label{ass1}
			The triple ($A_i, B_i, C_i$) is controllable and observable, and the graph $\mathcal{G}$ is an undirected graph.
		\end{assumption}	
	
		For HMAS (\ref{e1}), our purpose is to design a control input $u_i(t)$ such that $\lim _{t\rightarrow \infty}y_i\left( t \right) =y_i^*$, where $y_i^*$ is an optimal steady vector. The set containing all possible vector $y_i^*$ is defined as $\mathbb{S}_i \subset  \mathbb{R}^q$, which is referred to as the steady-state set. 
	
		\begin{problem}\label{pro1}
			For HMAS (\ref{e1}) with at most $r$-local/total Byzantine agents, the control objective is to devise a controller $u_i$ such that the output state $y_i$ of each normal agent converges to an optimal output state $y_i^*$ that solves the optimization problem:
			\begin{align}\label{e2a}
				&\min \sum\nolimits_{i=1}^{\mathcal{N}}{f_i\left( y_i \right)},\ y_i \in \mathbb{R}^{q}\\
				&\mathrm{s.t.} \ y_i=y_j, \forall i,j \in \mathcal{I_N}  \nonumber 
			\end{align}
			where  ${f_i\left( y_i \right)}: \mathbb{R}^{q} \rightarrow \mathbb{R}$ is a local cost function known only to agent $i$ itself.
		\end{problem}
		
		\begin{assumption}\label{ass2}
			Each normal agent is embedded with local cost function $f_i(\cdot)$ that is continuous differentiable, strongly convex and locally Lipschitz, i.e., $\forall a_1, a_2 \in \mathbb{R}^{q}$, $\left\| f_i\left( a_1 \right) -f_i\left( a_2 \right) \right\|^2 \leqslant \vartheta _i\left\| a_1-a_2 \right\|^2$ with $\vartheta _i$ being a positive constant.
		\end{assumption}
		
		Let $f\left( y \right) =\sum\nolimits_{i=1}^{\mathcal{N}}{f_i\left( y_i \right)}$, $y=\mathrm{col}\left\{ y_1,...,y_{\mathcal{N}} \right\}$, be the global cost function of all normal agents.

\begin{lemma}\textit{\cite{an2021distributed}} \label{lemma1}
	For system \eqref{e2b}, if the state of agent $i$ converges to a steady state $\mathrm{col}\{x_{i,1}^*, x_{i,2}^*\}$, it holds that
	$A_{i,11}x_{i,1}^{*}+A_{i,12}x_{i,2}^{*}=0$ and $y_i^* \in \mathbb{S}_i$,
	where the steady-state set  $\mathbb{S}_i$ satisfies $\mathbb{S}_i=C_i\mathrm{ker}(\left[ A_{i,11}\,\,A_{i,12} \right])$.
\end{lemma}
		
	   By Problem \ref{pro1} and Lemma \ref{lemma1}, we conclude that the optimal output state of agent $i$, $\forall i \in \mathcal{I_N}$, belongs to set $\mathbb{S}_i$. Since the constraint $y_i=y_j$ holds, it is essential to ensure that all agents share a nonempty common steady-state set to guarantee the existence of a solution for Problem \ref{pro1}. Consequently, the following assumption must be satisfied.
		\begin{assumption}\label{ass3}
			The solution $y^*\in \mathbb{R}^{\mathcal{N}q}$ of $\min_{y\in \mathbb{R} ^{\mathcal{N}q}}f\left( y \right)$ is within the intersection of steady-state sets of all agents, i.e., $y_i^*\in \bigcap_{i=1}^{\mathcal{N}}{\mathbb{S} _i}$, $\forall i \in \mathcal{I_N}$.
		\end{assumption}

		Under Assumption \ref{ass3}, formula (\ref{e2a}) can be reformulated to
		\begin{align}\label{e3a}
		&\underset{y_i\in \bigcap_{i=1}^{\mathcal{N}}{\mathbb{S} _i}}{\min}\sum\nolimits_{i=1}^{\mathcal{N}}{f_i\left( y_i \right)}
		\\
	 	&\mathrm{s.t.} \ y_i=y_j, [ A_{i,11}\,\,A_{i,12} ]y_i= 0, \forall i,j \in \mathcal{I_N}  \nonumber 
		\end{align}
		
\section{Edge-based Event-triggered Adaptive Distributed Optimization Algorithm}
In this section, a novel edge-based distributed optimization protocol for HMASs based on an adaptive hybrid event-triggered mechanism is constructed.

Define $t_{k_{ij}}^{ij}$, $k_{ij} \in \mathbb{Z}^+$, as the triggering instant from agent $j$ to agent $i$, in which  ${k_{ij}}$ denotes the corresponding sequence number of the trigger, determined by $k_{ij}=\mathrm{arg}\min _{\upsilon}\left\{ t-t_{\upsilon}^{ij}|t\geqslant t_{\upsilon}^{ij},\upsilon \in \mathbb{Z} ^+ \right\}$.  For convenience, in what follows, we refer to $t_k^{ij}$ as $t_{k_{ij}}^{ij}$ when there is no confusion, the notation ``$ij$" in  $( \cdot )^{ij}$ or $( \cdot ) _{ij}$ signifies the parameter transmitted from  agent $j$ to agent $i$.

 \subsection{Hybrid Adaptive Event-Triggered Protocol}		
Construct the closed-loop distributed control protocol $\eqref{e1a}$ for HMASs \eqref{e1} as 
		\begin{align}\label{e3}
			u_i=&-\hat{B}_i\left( \mathcal{D} _i\varUpsilon _i-\mathcal{S} _ix_i \right) 
			\nonumber\\
			\dot{\delta}_i=&-\rho C_{i}^{T}\triangledown f_i ( C_i \delta _{i}) -\alpha C_{i}^{T}\sum_{j=1}^{\mathcal{N}}{a_{ij}\hat c_{ij}( C_i\hat{\delta}_{ji}-C_j\hat{\delta}_{ij} )}
			\nonumber\\
			&-\beta C_i^T \sum_{j=1}^{\mathcal{N}}{a_{ij}\left( C_i \hat{w}_{ji}- C_j \hat{w}_{ij} \right)}
			\nonumber\\
			\dot{w}_i=&\alpha C_{i}^{T}\sum_{j=1}^{\mathcal{N}}{a_{ij}\hat c_{ij} ( C_i\hat{\delta}_{ji}-C_j\hat{\delta}_{ij})}\nonumber\\
			\dot{c}_{ij}=&\eta _{ij} \| C_i\hat{\delta}_{ji}-C_i\hat{\delta}_{ij} \| ^2
		\end{align}
		where $\forall j\in \mathcal{N} _i$, $t\in [ t_{{k}}^{ji},t_{{k}+1}^{ji})$, $\hat{\delta}_{ij}=\delta _{ij}(t_{k}^{ij}),\hat{w}_{ij}=w_{ij}(t_{k}^{ij})$,
	the gain parameters $\alpha, \beta, \rho>0$, $\eta_{ij}=\eta_{ji}>0$; the adaptive coupling weight $\hat{c}_{ij}=c_{ij}( t_{k}^{ij} )$, $\hat c_{ij}=\hat c_{ji}$ and $c_{ij}(0)\geqslant 1$; the matrices $\hat{B}_i=\bar{B}_{i}^{T}\left( \bar{B}_i\bar{B}_{i}^{T} \right) ^{-1}$, $\mathcal{D} _i=\left[F_i\bar{K}_i	\ \bar{K}_i\right]$, $\varUpsilon _i=\mathrm{col}\{ \delta _i, \dot \delta _i \}$,
		$\mathcal{S} _i=\bar{K}_iA_i+F_i\bar{K}_i$, $\bar{K}_i=\left[ K_i\,\,-I \right]$,
		$F_i=\bar{\mu}I+ A_{i,12}^{T}A_{i,12} $, with $K_i \in \mathbb{R} ^{p_i\times p_i}$ and $\bar{\mu} > 0$ to be designed.
		
		For the triggering instant $t_k^{ji}$ on path $\mathcal{P} \left( j,i \right)$, $\forall j \in {\mathcal{N}}_i$, the next triggering instant $t_{k+1}^{ji}$ is determined by the hybrid edge-based event-triggered condition (ETC):
		\begin{align}\label{e4}
			t_{k+1}^{ji}=\mathrm{inf}\Big\{&t>t_{k}^{ji}+T_{\rm MEI}^{ji}\big|f_1^{ji}(t)>0  \nonumber\\
			& \lor f_2^{ji}(t)>0 \lor f_3^{ji}(t)>0, T_{\rm MEI}^{ji}>0\Big\}
		\end{align}
		with
		$$\begin{aligned}
			f_1^{ji}(t)=&\hat c_{ji}\left\| C_ie_{{ji}}^{\delta} \right\| ^2-\gamma_\delta
			\\
			f_2^{ji}(t)=&\left\| C_ie_{{ji}}^{w} \right\| ^2-\gamma_w \\
			f_3^{ji}(t)=& c_{ji}- \hat c_{ji}-\gamma_c
		\end{aligned}$$
		where $e_{ij}^{\delta}=\hat{\delta}_{ij}-\delta _{ij},e_{ij}^{w}=\hat{w}_{ij}-w_{ij}$, and $\gamma_\delta, \gamma_w \in \mathbb{R}$ can be any function belonging to $\mathcal{L}_1$, and $\gamma_c$ is a positive value.
		Note that $t_{\rm MEI}^{i}$  represents the minimal event-triggered interval duration during which no trigger occurs. Specifically, in $[t_{k}^{ji}, t_{k}^{ji}+T_{\rm MEI}^{ji})$, the event-triggered mechanism of agent $i$ does not work, while in $[t_{k}^{ji}+T_{\rm MEI}^{ji}, t_{k+1}^{ji})$, the event-triggered mechanism is active. With the edge-based event-triggered mechanism, when a trigger occurs for agent $i$ on path $\mathcal{P}(j,i)$, the values $\hat \delta_{ji}$, $\hat w_{ji}$, $\hat c_{ji}$ will be immediately updated in \eqref{e3} for both agents $i$ and $j$.
		
\begin{figure}
	\centering
	\vspace{0.2cm}
	\subfigtopskip=1pt 
	\subfigbottomskip=1pt
	\subfigcapskip=2pt
	\includegraphics[width=0.4 \textwidth]{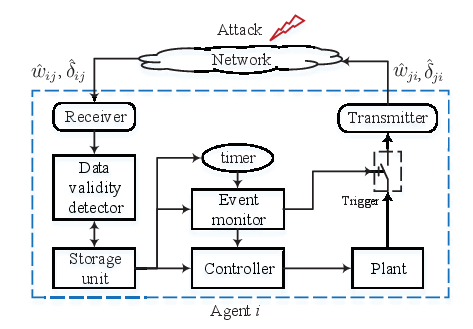}
	\caption{Self-triggered detection-based control approach for the HMASs.}\label{fig3} %(c) Hierarchical form of $\mathcal{G}$ with effective communication links.
\end{figure}

\begin{remark}
When agent $i$ executes the event-triggered mechanism in $[t_{k}^{ji}+T_{\rm MEI}^{ji}, t_{k+1}^{ji})$, the variables in ETC \eqref{e4} only utilize the stored neighboring information $\hat \delta_{ij}$ and  $\hat w_{ij}$, $j \in \mathcal{N}_i$. Accordingly, the coupling weight $\hat c_{ij}$, $j \in \mathcal{N}_i$, can be obtained by itself rather than by agent $j$. From ETC \eqref{e4}, one can observe that each agent determines when to trigger depending on its own stored information rather than continuously monitoring the neighbor's real-time information. This implies that the event-triggered mechanism is in essence a self-triggered mechanism.
Fig. \ref{fig3} demonstrates the schematic diagram of the self-triggered detection-based control approach with Byzantine agents.
\end{remark}

\subsection{Description of Activation Function}	
		We construct an activation function $m_{ji}$ as follows:
		\begin{align}\label{e16}
			\dot{m}_{ji}=&-\hat{s}_{ji}(\sigma _1m_{ij}+\sigma _2 )
			  \\
			\hat{s}_{ji}=&\left\{ \begin{array}{l}
				\begin{matrix}
					1,&		t\in [t_{k}^{ji},t_{k}^{ji}+T_{\rm MEI}^{ji})\\
				\end{matrix}\\
				\begin{matrix}
					0,&		t\in [t_{k}^{ji}+T_{\rm MEI}^{ji},t_{k+1}^{ji})\\
				\end{matrix}\\
			\end{array} \right. \nonumber
		\end{align}
		where ${m}_{ji} \in \mathbb{R}$, $\sigma _1$, $\sigma _2$  are positive constants related to $\hat c_{ji}$, which will be designed later; $m_{ji}(t_{k}^{ji+})=\frac{1}{\hat{c}_{ij}} m_{ji} (0)$ is set with $m_{ji} (0)$ being a positive value. The variable \(\hat{s}_{ji}\) acts as the control switch for \(m_{ji}\), determining when \(m_{ji}\) changes.
		
		In view of \eqref{e16}, one obtains $\dot m_{ji}<0$ in $[t_{k}^{ji},t_{k}^{ji}+T_{\rm MEI}^{ji})$, which indicates that $m_{ji}$ keeps monotonically decreasing in $[t_{k}^{ji},t_{k}^{ji}+T_{\rm MEI}^{ji})$. To guarantee the positivity of $m_{ji}$ and the existence of $T_{\rm MEI}^{ji}$, the following analysis is employed. 
	
From \eqref{e16}, we have that  $\forall t \in [t_{k}^{ji},t_{k}^{ji}+T_{\rm MEI}^{ji})$, 
\begin{align}\label{e15c}
m_{ji}\left( t \right) =e^{-\sigma _1\left( t-t_{k}^{ji} \right)}m_{ij} ( t_{k}^{ji} ) +\frac{\sigma _2}{\sigma _1}\left[1-e^{\sigma _2\left( t-t_{k}^{ji} \right)} \right] 
\end{align}
On the basis of equation \eqref{e15c}, there must exist a positive value $T_0$ such that ${m}_{ji}(t_{k}^{ji}+T_0)=0$, where
\begin{align}\label{e16a}
T_0=\frac{1}{\sigma _1}\ln \left( \frac{1}{2}+\frac{1}{2}\sqrt{1+\frac{4\sigma _1}{\sigma _2}m_{ji}(  t_{k}^{ji+})} \right) 
\end{align}
If the minimal event-triggered interval $T_{\rm MEI}^{ji}$ is set such that $T_{\rm MEI}^{ji}<T_0$, the positivity of ${m}_{ji}$ can be maintained for $t\in [t_{k}^{ji},t_{k}^{ji}+T_{\rm MEI}^{ji})$. Furthermore, since $\hat{s}_{ji}$ is set to be zero, the parameter ${m}_{ji}$ is fixed in $[t_{k}^{ji}+T_{\rm MEI}^{ji},t_{k+1}^{ji})$.
This implies that $\forall t>0$, $m_{ji} >0$ always holds.

\begin{remark}
	The activation function $m_{ji}$ is utilized to guarantee the presence of the positive MEI among agents, which will be applied to the subsequent attack detection approach.  It is crucial to note that the positivity of $m_{ji}$ must be satisfied, as it is a variable that will be incorporated into the Lyapunov function in Section \ref{sectionV}.
\end{remark}

\section{Attack Detection and Isolation}
In this section, we propose a threshold-based detection and isolation algorithm for each agent under attacks. Before illustrating the main algorithm, we present a discussion to help readers understand the threshold-based detection approaches. Until now, there are several threshold-based detection methods to identify the potential Byzantine agents in networks, including the approaches based on performance metrics \cite{li2019performance}, voting mechanisms \cite{Xie2019}, and machine learning \cite{Blanchard2017}. However, none of these can completely detect all  types of Byzantine agents. In fact, there is no perfect threshold-based detection approach for Byzantine attacks across all scenarios. This limitation highlights the need for a new threshold-based detection approach to more effectively address these challenges.

Similar to \cite{zegers2021event}, \cite{li2019performance,Xie2019,Blanchard2017}, the proposed threshold-based approach in this paper cannot handle the Byzantine agents who fully comprehend the detection strategy. Despite this, the self-triggered hybrid detection approach introduced here demonstrates applicability in specific scenarios, offering substantial technical support for Byzantine detection techniques.

\begin{figure}
	\centering
	\vspace{0.2cm}
	\subfigtopskip=1pt 
	\subfigbottomskip=1pt
	\subfigcapskip=2pt
	\subfigure[]{\includegraphics[width=0.2 \textwidth]{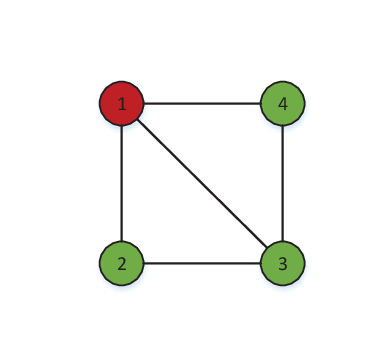}}
	\subfigure[]{\includegraphics[width=0.2 \textwidth]{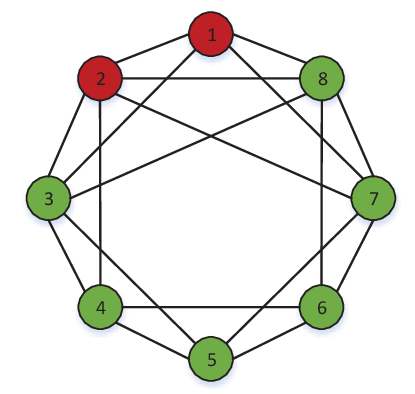}}
	\caption{ (a) $1$-isolatable graph. (b) $2$-isolatable graph. }\label{fig1} 
\end{figure}

\subsection{$\mathit{r}$-Isolatable Graph}
\begin{definition}\textit{($(r,s)$-connected graph)}
	An undirected graph $\mathcal{G}$ is referred to as an $r$-connected graph, if for any nonempty disjoint sets $\mathcal{X}_1, \mathcal{X}_2 \subset \mathcal{V}$, one of the following conditions holds:
	$$\begin{aligned}
		(1)\ \mathfrak{D} _{\mathcal{X} _1}^{r} =\left| \mathcal{X} _1 \right|;
		(2)\  \mathfrak{D} _{\mathcal{X} _2}^{r} =\left| \mathcal{X} _2 \right|;
		(3)\  \mathfrak{D} _{\mathcal{X} _1}^{r} + \mathfrak{D} _{\mathcal{X} _2}^{r} \geqslant s.
	\end{aligned}$$
	where $s \in \mathbb{Z}^+$, $\mathfrak{D} _{\mathcal{X} _p}^{r}$, $p=1,2$, represents the cardinality of nodes in $\mathcal{X}_p$ that each have least $r$ neighbors out of $\mathcal{X}_p$. Specifically,  $\mathfrak{D} _{\mathcal{X} _p}^{r}=| \mathcal{Z}_{\mathcal{X} _p}^{r} |$, with
	$\mathcal{Z} _{\mathcal{X} _p}^{r}=\left\{ i\in \mathcal{X} _p\left| | \mathcal{N} _i\setminus \mathcal{X} _p |\geqslant r \right. \right\}$.
\end{definition}

To simplify terminology, an $(r, 1)$-connected graph will be referred to as an  $r$-connected graph.
\begin{definition}\textit{($r$-isolatable)}
	In undirected graph $\mathcal{G}$, if any $r$ agents are removed (i.e. isolated) by their neighbors, the remaining graph $\mathcal{G_N}$ is still connected, then $\mathcal{G}$ is termed as $r$-isolatable.
\end{definition}

\begin{lemma}\label{lemma.2}
	If graph $\mathcal{G}$ is an $(r+1)$-connected graph, then the graph must be an $r$-isolatable graph.
\end{lemma}

\emph{Proof:} See Section \ref{appendixA}. $\hfill\blacksquare$

\begin{remark}
	We illustrate two cases in Fig. \ref{fig1} to describe the concept of $r$-isolatable graph. Note that when any agent in Fig. \ref{fig1} is isolated by all its neighbors, the remaining graph $\mathcal{G_N}$ is still connected. This implies that the graph depicted in Fig. \ref{fig1} (a) is $1$-isolatable. In contrast, the graph in Fig. \ref{fig1} (b) is $2$-isolatable.
\end{remark}

\subsection{Attack Detection and Isolation Mechanism}
\begin{figure}[!]
	\centering
	\vspace{0.2cm}
	\subfigtopskip=1pt 
	\subfigbottomskip=1pt
	\subfigcapskip=2pt
{\includegraphics[width=0.36 \textwidth]{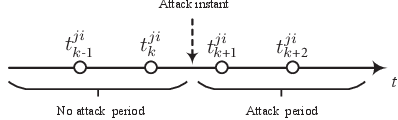}}
	\caption{Information of launching the attack in a trigger interval.}\label{fig2} 
\end{figure}

During the triggering interval $[ t_{k}^{ij},t_{k+1}^{ij})$, for agent $i$, since $\hat \delta_{ij}$ and $\hat w_{ij}$ are two transmission variables, their true errors (which are not tampered with) between instants $t_{{k}}^{ij}$ and $t_{{k}+1}^{ij}$ are represented as follows:
\begin{align}
\bar e_{ij}^{\delta}(t_{k+1}^{ij})=&\delta_j(t_{k}^{ij})- \delta_j(t_{k+1}^{ij})\\
\bar e_{ij}^{w}(t_{k+1}^{ij})=& w_j(t_{k}^{ij})-w_j(t_{k+1}^{ij})
\end{align}
However, we know that their actual errors between instants $t_{{k}}^{ij}$ and $t_{{k}+1}^{ij}$ are
\begin{align}
	e_{ij}^{\delta}(t_{k+1}^{ij})=& \delta _{ij} (t_{k}^{ij})-{\delta}_{ij} (t_{k+1}^{ij}) \label{e14a}\\
	e_{ij}^{w}(t_{k+1}^{ij})=& w_{ij} (t_{k}^{ij})-w_{ij} (t_{k+1}^{ij})\label{e14b}
\end{align}
At each trigger instant $t_{k+1}^{ij}$, $k \in \mathbb{Z}^+$, the true values $\delta _{j} (t_{k+1}^{ji})$ and $ w_{j} (t_{k+1}^{ji})$ of agent $j$ is unknown for agent $i$. However, if agent $j$ is a normal agent, the conditions $e_{ij}^{\delta}(t_{ k+1}^{ij})= \bar e_{{ij}}^{\delta}(t_{ k+1}^{ij})$ and $e_{{ij}}^{w}(t_{  k+1}^{ij})=\bar e_{{ij}}^{w}(t_{ k+1}^{ij})$ are holds. In contrast, when agent $j$ behaves abnormal (i.e., Byzantine agents), at least one of the conditions \eqref{e14a} and \eqref{e14b} will be invalid. Fig. \ref{fig2} provides a schematic diagram of the attacker launching an attack during the triggering interval. It indicates that when the Byzantine agent tampers with its values in $[ t_{k}^{ij},t_{k+1}^{ij})$, the error $e_{ij}^{\delta}(t_{k+1}^{ij})$ and $e_{ij}^{w}(t_{k+1}^{ij})$ are likely to undergo a large change.

By leveraging this characteristic, we choose the time-varying thresholds $\digamma_{ij}^{\delta}$ and $\digamma_{ij}^{w}$ such that the conditions $\gamma _{\delta}\leqslant \digamma _{ij}^{\delta},\gamma _w\leqslant \digamma _{ij}^{w}$ holds. This setting thereby ensures the following inequalities:
\begin{align}
	c_{ij}(t_{k}^{ij})\| C_je_{ij}^{\delta}(t_{k}^{ij})\|^2\leqslant& \digamma _{ij}^{\delta},\label{e15a}\\
	\| C_je_{ij}^{w}(t_{k}^{ij})\|^2\leqslant& \digamma _{ij}^{w} \label{e15b}
\end{align}
Consequently, since at least one of the above conditions may be invalid if agent $j$ misbehaves, we obtain that if either $c_{ij}( t_{ k}^{ij} )\| C_j {e}_{{ij}}^{\delta}( t_{ k}^{ij} )\|^2 \leqslant \digamma _{ij}^{\delta}$ or $\| C_j {e}_{{ij}}^{w} ( t_{ k}^{ij}) \|^2 \leqslant \digamma _{ij}^{w}$ are not true, agent $j$ is regarded as a Byzantine agent and should be isolated immediately by agent $i$. 

Moreover, under the hybrid ETC \eqref{e4}, the interval $t_{k+1}^{ij}-t_{k}^{ij}$  of two consecutive triggers is not smaller than $T_{\rm MEI}^{ij}$ if agent $j$ is deemed normal. In other words, for agent $i$, once $t_{k+1}^{ij}-t_{k}^{ij}<T_{\rm MEI}^{ij}$, agent $j$ will be identified as the Byzantine agent and need to be isolated at once.

For normal agent $i$, if neighbor $j$ is to be isolated at time $t_1 > 0$, we will set $a_{ij}(t) = 0$ for all $t \geqslant t_1$. Note that isolating any neighbor effectively alters the Laplacian matrix of the communication graph. However, such isolation decision is kept private, known only to the agent $i$ and not shared with its neighbors, resulting in that the actual Laplacian matrix is unknown for all agents. By utilizing the adaptive protocol \eqref{e3}, this issue of hidden changes in network topology can be solved. The detailed procedure of Byzantine attacks detection and isolation (BADI) algorithm is presented in Algorithm \ref{alg:1}. 

Note that, in this paper, we assume that Byzantine agents cannot intelligently manipulate data to evade detection by the BADI algorithm, and all normal agents using BADI algorithm can effectively detect and isolate all Byzantine agents. This assumption is practical because, in some occasions, Byzantine attackers often lack detailed knowledge about communication  details (such as trigger intervals and detection thresholds), which can lead them to inadvertently send detectable data, especially when attempting to disrupt the system quickly.

\begin{remark}
	One can select small values for $\gamma _{\delta}$ and $\gamma _{\delta}$ in ETC \eqref{e4}, and then choose small thresholds $\digamma _{ij}^{\delta}$, $\digamma _{ij}^{w}$ to ensure that $ \digamma _{ij}^{\delta} \geqslant \gamma _{\delta}$, $\digamma _{ij}^{w} \geqslant \gamma _{w}$. Under this setting, the BADI algorithm can more accurately detect the Byzantine agents if the transmitted values have been subtly tampered with. In addition, because the proposed BADI algorithm is able to detect the abnormal data transmission between any two agents, it can not only be applied to the detection of Byzantine attacks on nodes, but also to other attacks on communication channels, such as replay attacks \cite{naha2022sequential}, false data injection attacks \cite{guo2023hybrid}.
\end{remark}
\begin{remark}
	In \cite{zegers2021event}, a trigger error threshold approach is employed for attack detection, but it allows the Byzantine agents to generate numerous triggers within a short period, with only minor parameter errors between consecutive triggers. This could enable the Byzantine agent to evade detection and cause significant deviations in parameters. In contrast, the approach proposed in this article introduces a known MEI, which effectively limits the number of triggers an agent can generate in a finite time. As a result, our method offers enhanced precision in detecting attacks compared to \cite{zegers2021event}.
\end{remark}

\begin{algorithm}[t!]
	\caption{BADI Algorithm}
	\label{alg:1}
	\begin{algorithmic}[1]\small
		\STATE Select appropriate thresholds $\digamma_{ij}^\delta$ and $\digamma_{ij}^w$.
		\FOR{$i = 1$ to ${\mathcal{N}}$, $t \geqslant 0$}
		\IF{a trigger occurs on $\mathcal{P} \left( j,i \right) $}
		\STATE Set $T_{\rm {MEI}}^{ij}$ according to \eqref{e16a}.
		\STATE Check the following conditions:\\
		\textbf{1) Trigger Interval Condition (TIC):} 
		$$t_{k+1}^{ij} - t_k^{ij} \geqslant T_{\rm {MEI}}^{ij}$$
		\textbf{2) Trigger Error Conditions (TEC):} 
		$$c_{ij}(t_k^{ij})\|C_j e_i^\delta(t_k^{ij})\|^2 \leqslant \digamma_{ij}^\delta, \|C_j e_{ij}^w(t_k^{ij})\|^2 \leqslant \digamma_{ij}^w$$
		\IF{both conditions are satisfied}
		\STATE Agent $j$ remains available for agent $i$.
		\ELSE
		\STATE Agent $j$ is marked as unavailable for agent $i$. Set $a_{ij} \gets 0$ and remove $j$ from $\mathcal{N}_i$ to isolate it.
		\ENDIF
		\ENDIF
		\ENDFOR
	\end{algorithmic}
\end{algorithm}

		\section{Convergence analysis}	\label{sectionV}
		\subsection{Convergence of Optimal Observer}
	Define $\{t_\varepsilon^r\}_{\varepsilon =1}^{\hbar }$, $\varepsilon \in \mathbb{Z} ^+$, as the time sequence of the instants when some agents are isolated by all their neighbors rather than some of their neighbors. The term $t_{\hbar}^{r}$, where $\hbar \leqslant \mathcal{Q}$ is a positive integer, denotes the instant when all Byzantine agents are isolated by their neighbors.
	
	Taking the consideration of Byzantine agents among the graph, one obtains that in time interval $[t_{\varepsilon}^r, t_{\varepsilon +1}^r)$,  
	\begin{align}\label{e9}
		\dot{x}_i=&A_ix_i-B_i\hat{B}_i\left( \mathcal{D} _i\varUpsilon _i-\mathcal{S} _ix_i \right) 
		\nonumber\\
		\dot{\delta}_{i}=&-\rho C_{i}^{T}\triangledown f_i(C_i {\delta}_{i})-\alpha C_{i}^{T}\sum_{j=1}^{\mathcal{N}}{a_{ij}\hat c_{ij}(C_i\hat{\delta}_{ji}-C_j\hat{\delta}_{ij})}
		\nonumber\\
		&-\beta C_{i}^{T}\sum_{j=1}^{\mathcal{N}}{a_{ij}\left( C_i\hat{w}_{ji}-C_j\hat{w}_{ij} \right)}
		\nonumber\\
		\dot{w}_{i}=&\alpha C_{i}^{T}\sum_{j=1}^{\mathcal{N}}{a_{ij}\hat c_{ij}(C_i\hat{\delta}_{ji}-C_j\hat{\delta}_{ij})}
		\nonumber\\
		\dot{c}_{ij}=&\eta _{ij}\|C_i\hat{\delta}_{ji}-C_j\hat{\delta}_{ij}\|^2
	\end{align}	
		
Denote $\delta _{i}^{*}$ and $w _{i}^{*}$ as the equilibrium points of $\delta _i$ and $w_i$, respectively, when all Byzantine agents are isolated from agent $i$. Then let $\tilde{\delta}_i(t)=\delta _i-\delta _{i}^{*}, \tilde{w}_i =w _i-w _{i}^{*}$. From \eqref{e9}, we have
		\begin{align}\label{e10}
			\dot{\tilde{\delta}}_i=&-\rho C_{i}^{T}\triangledown f_i(C_i\delta _i)
			\nonumber\\
			&-\alpha C_{i}^{T}\sum_{j=1}^{\mathcal{N}}{a_{ij}\hat{c}_{ij}\left[ C_i( \tilde{\delta}_{ji}+e_{ji}^{\delta} ) -C_j( \tilde{\delta}_{ij}+e_{ij}^{\delta}) \right]}
			\nonumber\\
			&-\beta C_{i}^{T}\sum_{j=1}^{\mathcal{N}}{a_{ij}\left[ C_i( \tilde{w}_{ji}+e_{ji}^{w} ) -C_j( \tilde{w}_{ij}+e_{ij}^{w}) \right]}+h_{i}^{\delta}
			\nonumber\\
			\dot{\tilde{w}}_i=&\alpha C_{i}^{T}\sum_{j=1}^{\mathcal{N}}{a_{ij}\hat{c}_{ij}\left[ C_i( \tilde{\delta}_{ji}+e_{ji}^{\delta}) -C_j( \tilde{\delta}_{ij}+e_{ij}^{\delta}) \right]}+h_i^w
		\end{align}
		where $h_{i}^{\delta}=-h_{i}^{w}-\beta C_{i}^{T}\sum_{j=1}^{\mathcal{N}}{}a_{ij}\left( C_i\hat{\epsilon}_{ji}^{w}-C_j\hat{\epsilon}_{ji}^{w} \right) ,h_{i}^{w}=\alpha C_{i}^{T}\sum_{j=1}^{\mathcal{N}}{}a_{ij}\hat{c}_{ij}(C_i\hat{\epsilon}_{ji}^{\delta}-C_j\hat{\epsilon}_{ij}^{\delta})$ with  $\hat{\epsilon}_{ji}^{\delta}=\epsilon _{ji}^{w}( t_{k}^{ji})$, $\hat{\epsilon}_{ji}^{w}=\epsilon _{ji}^{w}( t_{k}^{ji})$.
		
	%For subsequent use, define the following column vectors and matrices:
	Then define	${\delta}=\mathrm{col}\{ {\delta}_1,..., {\delta}_{\mathcal{N}}\}$, $\delta ^*=\mathrm{col}\{\delta _{1}^{*},...,\delta _{\mathcal{N}}^{*}\}$, $\hat{\delta}=\mathrm{col}\{ \hat{\delta}_1,...,\hat{\delta}_{\mathcal{N}}\}$, $w =\mathrm{col}\left\{ w _1,...,w _{\mathcal{N}} \right\}$, $w ^*=\mathrm{col}\left\{ w _{1}^{*},...,w _{\mathcal{N}}^{*} \right\}$, $\hat{w}=\mathrm{col}\{ \hat{w}_1,...,\hat{w}_{\mathcal{N}} \}$, $e^{\delta}=\mathrm{col}\{ e_{1}^{\delta},...,e_{{\mathcal{N}}}^{\delta} \}$, $e^{w}=\mathrm{col}\{ e_{1}^{w},...,e_{{\mathcal{N}}}^{w}\}$, $C=\mathrm{diag}\{ C_1,...,C_{{\mathcal{N}}}\}$, $h^{\delta}=\mathrm{col}\{ h_{1}^{\delta},...,h_
		\mathcal{N}^{\delta}\}$, $h^w=\mathrm{col}\{ h_{1}^{w},...,h_\mathcal{N}^{w}\}$.
		
		\begin{lemma}\label{lamma6}
			Assume that Assumption \ref{ass1}-\ref{ass3} hold and graph $\mathcal{G}$ is an $(r+1)$-connected graph. Let $( \delta_i^*,  w_i^*)$ be one of its equilibrium points of \eqref{e9}. Then $C_i\delta_i^*$ is one of the optimal solutions of Problem \ref{pro1} with $C_i\delta _{i}^{*}=C_j\delta _{j}^{*},\forall i,j\in \mathcal{I} _{\mathcal{N}}$, if all Byzantine agents are isolated from the normal agents.
		\end{lemma}
		
		\emph{Proof}: At the equilibrium point $( \delta_i^*,  w_i^*)$, it holds $e_{ij}^{\delta}=e_{ij}^{w}=0$.  Since all Byzantine agents are isolated, we have $h^{\delta}=h^w=0$. It follows from \eqref{e10} that
		\begin{align}
			0=&-\rho C^T\triangledown f(C \delta^*)-\alpha C^T(\widehat{L}\otimes I)C \delta^*\nonumber\\
			&-\beta C^T(L\otimes I)C w^*
			\label{e11d} \\
			0=&\alpha C^T(\widehat{L}\otimes I)C \delta^* \label{e11}
		\end{align}
where $\widehat{L}=\left[\hat c_{ij}L_{ij} \right] \in \mathbb{R}^{\mathcal{N} \times \mathcal{N}}$ is symmetric. Since $\mathcal{G}$ is an $(r+1)$-connected graph, from Lemma \ref{lemma.2}, we have that $\mathcal{G}$ is $r$-isolatable. Therefore, there must exist a vector $\delta _c\triangleq \mathbf{1} \otimes \bar{\varsigma}$, $\forall \bar{\varsigma}\in \mathbb{R} ^q$, satisfying $(\widehat{L}\otimes I)\delta _c=0$. By \eqref{e11d} and \eqref{e11}, one has 
		\begin{align}
			C_i{\delta}_i^*=&\bar{\varsigma},\  \forall i=1,..,\mathcal{N} \label{e11a}\\
			\rho C^T\triangledown f(C \delta^*)=&-\beta C^T(L\otimes I)C{w}^*  \label{e11b}
		\end{align}
		Multiplying $C$ left on both sides of equation \eqref{e11b} yields $\rho C C^T \triangledown f(C\delta ^*)=-\beta C C^T (L\otimes I)Cw^*$. Because $C$ is of full column rank,  $C C^T$ is positive definite, which implies $\rho \triangledown f(C\delta ^*)=-\beta (L\otimes I)Cw^*$. With the fact that $\mathbf{1}^TL=0$, we have $\rho \sum\nolimits_{i=1}^{\mathcal{N}}{\triangledown f(C_{i}\delta_i^*)}=\beta (\mathbf{1}^{T}L\otimes I)C {w}^*=0$. Since the function $f(\cdot)$ is strongly convex, it follows from $\sum\nolimits_{i=1}^{\mathcal{N}}{C_{i}^{T}\triangledown f(C_i {\delta}_i^*)}=0$ that point $ C_i\delta_i^*$ is a global optimal solution of \eqref{e2a}, in which $C_i\delta _{i}^{*}=C_j \delta _{j}^{*},\forall i, j=1,...,\mathcal{N}$.  $\hfill\blacksquare$
		
		\begin{theorem}\label{theoreom.1a}
			Suppose Assumptions \ref{ass1}-\ref{ass3} are well-posed and graph $\mathcal{G}$ is an $(r+1)$-connected graph. For HMAS \eqref{e1} with $r$-local/total Byzantine agents, the protocol \eqref{e3} under Algorithm \ref{alg:1} can ensure that $\lim _{t\rightarrow \infty}\delta _i=\delta _{i}^{*}$, $\lim _{t\rightarrow \infty}w_i=w_{i}^{*}$ if there exists an instant such that all Byzantine agents are isolated by their normal neighbors, and the variables $\alpha, \beta, \rho>0$, 
			\begin{align}
			 \sigma _1 = &\lambda _{M}^{2}( C) \hat{c}_{ij}{b}_M  	\label{e12a}
			 \\
			 \sigma _2=&\frac{1}{2}\lambda _{M}^{2}( C ) \hat{c}_{ij}\alpha \phi	\label{e12a2}
			\end{align}
		where  ${b}_M=\max \left\{ b_1,b_2 \right\}$, $\phi >\max \{ \frac{2b_3}{\alpha},\frac{5\beta}{2\alpha}\}$, $b_1=3\alpha +\frac{5}{2}\beta +\frac{2\bar{\eta}\gamma _c}{d_M}+\frac{\rho}{2d_M}$, $b_2=\alpha +\beta$, $b_3=(2\alpha +\frac{2}{d_M}\bar{\eta}\gamma ^c)\bar{m}$, $\bar{\eta}=\max_{i\in \mathcal{I} _{\mathcal{N}},j\in \mathcal{N} _i} \left\{ \eta _{ij} \right\}$, $\bar{m}=\max_{i\in \mathcal{I} _{\mathcal{N}},j\in \mathcal{N} _i} \left\{ m_{ij}\left( 0 \right) \right\}$. The $T_{\rm MEI}^{ji}$ in each successive triggering interval is strictly positive.
		\end{theorem}
		
		\emph{Proof}: 
	Select the Lyapunov candidate function in time interval $[t_{\varepsilon}^r, t_{\varepsilon+1}^r)$ as $V=V_1+V_2$, with
\begin{align}
V_1=&\frac{1}{2}\phi \sum_{i=1}^{\mathcal{N}}{\tilde{\delta}_{i}^{T}\tilde{\delta}_i}+\frac{1}{2}\sum_{i=1}^{\mathcal{N}}{(\tilde{\delta}_i+\tilde{w}_i)^T(\tilde{\delta}_i+\tilde{w}_i)}
\nonumber\\
&+\frac{1}{2}\phi _c\sum_{i=1}^{\mathcal{N}}{\sum_{j=1}^{\mathcal{N}}{a_{ij}\frac{1}{\eta _{ij}}}}(c_{ij}- \bar c )^2
\label{e25th_a}\\
V_2=&\frac{1}{d_M}\sum_{i=1}^{\mathcal{N}}{\sum_{j=1}^{\mathcal{N}}{a_{ij}}}m_{ij}\left(\| e_{ji}^{\delta}\| ^2+\| e_{ji}^{w}\|^2+ \|e_{ji}^{c}\|^2 \right) \label{e25th_b}
\end{align}
where $\phi _{c}=\frac{1}{4}\alpha \phi -\frac{1}{2}b_3>0$, and $\bar c$ is arbitrary positive value. %Note that $\frac{\partial}{\partial t}(e_{ji}^{\delta}+\tilde{\delta}_i)=\frac{\partial}{\partial t}(\hat{\delta}_{ji}-\delta ^*)=0,\frac{\partial}{\partial t}(e_{ji}^{w}+\tilde{w}_i)=\frac{\partial}{\partial t}(\hat{w}_{ji}-w^*)=0$.

Tacking the set-valued Lie derivative of $V_1$ in \eqref{e25th_a} along the trajectory of \eqref{e10} yields
		$$\begin{aligned}
\ell \dot{V}_1=&\mathbb{F} \Big[-\rho \left( \phi +1 \right) \sum_{i=1}^{\mathcal{N}}{\tilde{\delta}_{i}^{T}C_{i}^{T}\mathbf{f}_i}
-\alpha \phi \tilde{\delta}^TC^T(\widehat{L}\otimes I)C\tilde{\delta}
\\
&-\alpha \phi \sum_{i=1}^{\mathcal{N}}{\sum_{j=1}^{\mathcal{N}}{\hat c_{ij}a_{ij}\tilde{\delta}_{i}^{T}C_{i}^{T}\left( C_ie_{ji}^{\delta}-C_je_{ij}^{\delta} \right)}}
\\
&-\beta \left( \phi +1 \right) \tilde{\delta}^TC^T(L\otimes I)C\tilde{w}
\\
&-\beta \left( \phi +1 \right) \sum_{i=1}^{\mathcal{N}}{\sum_{j=1}^{\mathcal{N}}{a_{ij}\tilde{\delta}_{i}^{T}C_{i}^{T}\left( C_ie_{ji}^{w}-C_je_{ij}^{w} \right)}}
\\
&-\rho \sum_{i=1}^{\mathcal{N}}{\tilde{w}_{i}^{T}C_{i}^{T}\mathbf{f}_i} - \beta \tilde{w}^TC^T(L\otimes I)C\tilde{w}
\\
&-\beta \sum_{i=1}^{\mathcal{N}}{\sum_{i=1}^{\mathcal{N}}{a_{ij}\tilde{w}_{i}^{T}C_{i}^{T}\left( C_ie_{ji}^{w}-C_je_{ij}^{w} \right)}}
\\
&+\phi _{c} \sum_{i=1}^{\mathcal{N}}{\sum_{j=1}^{\mathcal{N}}{a_{ij}\left( c_{ij}-\bar c \right)}\| C_i\tilde{\delta}_i-C_j\tilde{\delta}_j\| ^2+\mathcal{H}}\Big]
		\end{aligned} $$
where  $\mathbf{f}_i=\triangledown f(C_i\delta _i)$, $\mathcal{H}=\phi \tilde{\delta}^Th^{\delta}+\tilde{\delta}^T\left( h^{\delta}+h^w \right) +\tilde{w}^T\left( h^{\delta}+h^w \right)$.
		
Let $\bar \phi= \phi+1$, $\vartheta = \max_{i=1,..,\mathcal{N}} \vartheta _i$.
From Assumption \ref{ass2}, we have $\left\| \mathbf{f}_i \right\| ^2 \leqslant \vartheta \| C_i \tilde\delta_i\| ^2$. Using the famous Young's inequality theorem, we have
$
-\rho \left( \phi +1 \right) \tilde{\delta}_{i}^{T}C_{i}^{T}\mathbf{f}_i\leqslant -\vartheta \rho \bar{\phi}\| C_i\tilde{\delta}_i\| ^2$
and
$
-\rho \tilde{w}_{i}^{T}C_{i}^{T}\mathbf{f}_i\leqslant \frac{1}{16}\beta \lambda _2\left\| C_i\tilde{w}_i \right\| ^2+\frac{4}{\beta \lambda _2}\vartheta \rho ^2\| C_i\tilde{\delta}_i\|^2$.
By letting $b_4=\frac{4}{\beta \lambda _{2}^{2}}\vartheta \rho ^2-\frac{ \bar{\phi}}{\lambda _2}\vartheta \rho$, one can verify
$$\begin{aligned}	
-\rho& \sum_{i=1}^{\mathcal{N}}{ [ \left( \phi +1 \right) C_i\tilde{\delta}_i+C_i\tilde{w}_i ]^T \mathbf{f}_i}
\\
\leqslant& b_4\tilde{\delta}^TC^T\left( L\otimes I \right) C\tilde{\delta}+\frac{1}{16}\beta \tilde{w}^TC^T\left( L\otimes I \right) C\tilde{w}
\end{aligned}$$
With the condition $\frac{1}{2}\sum_{i=1}^{\mathcal{N}}{\sum_{j=1}^{\mathcal{N}}{a_{ij}\hat{c}_{ij}\| C_i\tilde{\delta}_i-C_j\tilde{\delta}_j\|^2}}=\tilde{\delta}^TC^T(\widehat{L}\otimes I)C\tilde{\delta}$, it further gets
		 $$\begin{aligned}	
&\alpha \phi \sum_{i=1}^{\mathcal{N}}{\sum_{j=1}^{\mathcal{N}}{\hat{c}_{ij}a_{ij}\tilde{\delta}_{i}^{T}C_{i}^{T}\left( C_ie_{ji}^{\delta}-C_je_{ij}^{\delta} \right)}}
\\
=&\frac{1}{2}\alpha \phi \sum_{i=1}^{\mathcal{N}}{\sum_{j=1}^{\mathcal{N}}{\hat{c}_{ij}a_{ij}\left( C_i\tilde{\delta}_i-C_j\tilde{\delta}_j \right)^T\left( C_ie_{ji}^{\delta}-C_je_{ij}^{\delta} \right)}}
\\
\leqslant& \frac{1}{4}\alpha \phi \sum_{i=1}^{\mathcal{N}}{\sum_{j=1}^{\mathcal{N}}{\hat{c}_{ij}a_{ij}\| C_i\tilde{\delta}_i-C_j\tilde{\delta}_j \| ^2}}
\\
&+\frac{1}{4}\alpha \phi \sum_{i=1}^{\mathcal{N}}{\sum_{j=1}^{\mathcal{N}}{\hat{c}_{ij}a_{ij}\left\| C_ie_{ji}^{\delta}-C_je_{ij}^{\delta} \right\| ^2}}
\\
\leqslant& \frac{1}{2}\alpha \phi \tilde{\delta}^TC^T(\widehat{L}\otimes I)C\tilde{\delta}+\frac{1}{2}\alpha \phi \sum_{i=1}^{\mathcal{N}}{\sum_{j=1}^{\mathcal{N}}{\hat{c}_{ij}a_{ij}\left\| C_ie_{ji}^{\delta} \right\| ^2}}
			\end{aligned} $$	
Through the similar operation above, one can verify
		$$\begin{aligned}
	-\beta	& \left( \phi +1 \right) \tilde{\delta}^TC^T(L\otimes I)C\tilde{w}
		\\
		&\leqslant 4\beta \bar{\phi}^2\tilde{\delta}^TC^T(L\otimes I)C\tilde{\delta}+\frac{1}{16}\beta \tilde{w}^TC^T(L\otimes I)C\tilde{w},
		\end{aligned} $$
	$$\begin{aligned}
	-\beta & \left( \phi +1 \right) \sum_{i=1}^{\mathcal{N}}{\sum_{j=1}^{\mathcal{N}}{a_{ij}\tilde{\delta}_{i}^{T}C_{i}^{T}\left( C_ie_{ji}^{w}-C_je_{ij}^{w} \right)}}
	\\
	&\leqslant \beta \bar{\phi}^2\tilde{\delta}^TC^T(L\otimes I)C\tilde{\delta}+\frac{1}{4}\beta \sum_{i=1}^{\mathcal{N}}{\sum_{j=1}^{\mathcal{N}}{a_{ij}\left\| C_ie_{ji}^{w} \right\| ^2}},
		\end{aligned} $$ and
		$$\begin{aligned}
-\beta &\sum_{i=1}^{\mathcal{N}}{\sum_{i=1}^{\mathcal{N}}{a_{ij}\tilde{w}_{i}^{T}C_{i}^{T}\left( C_ie_{ji}^{w}-C_je_{ij}^{w} \right)}}
\\
&\leqslant \frac{1}{4}\beta  \tilde{w}^TC^T(L\otimes I)C\tilde{w}+ \beta \sum_{i=1}^{\mathcal{N}}{\sum_{j=1}^{\mathcal{N}}{a_{ij}\left\| C_ie_{ji}^{w} \right\| ^2}}
	\end{aligned} $$
In addition, the term $\mathcal{H}$ can be expressed as
$\mathcal{H} \leqslant \| \tilde{\delta}\| \bar{h} +\| \tilde{w} \| \bar{h}
	\leqslant \frac{1}{4}\alpha \tilde{\delta}^TC^T\left( L\otimes I \right) C\tilde{\delta}+ \frac{1}{8} \beta \tilde{w}^TC^T\left( L\otimes I \right) C\tilde{w}+\bar{\mathcal{H}}$
where $\bar{h}=\max \left\{ \| \bar \phi h^{\delta}+h^w \|, \| h^{\delta}+h^w \| \right\}$ and $\bar{\mathcal{H}}=\left(\frac{1}{\alpha} +\frac{2}{\beta} \right) \lambda _2\lambda _m^2\left( C \right)\bar{h}^2$.
As a result, we have
\begin{align}\label{e23}
\ell \dot{V}_1&\leqslant-\frac{1}{2}\alpha \phi \tilde{\delta}^TC^T(\widehat{L}\otimes I)C\tilde{\delta}
	\nonumber	\\
&+\phi _c\sum_{i=1}^{\mathcal{N}}{\sum_{j=1}^{\mathcal{N}}{a_{ij}\left( c_{ij}-\bar c \right)}\| C_i\tilde{\delta}_i-C_j\tilde{\delta}_j\| ^2}\nonumber \\
&+\left(b_4+5\beta \bar{\phi}^2+\frac{1}{4}\alpha\right)\tilde{\delta}^TC^T(L\otimes I)C\tilde{\delta} \nonumber\\
&-\frac{1}{2}\beta \tilde{w}^TC^T(L\otimes I)C\tilde{w}
+\frac{1}{2}\alpha \phi \sum_{i=1}^{\mathcal{N}}{\sum_{j=1}^{\mathcal{N}}{\hat{c}_{ij}a_{ij}\left\| C_ie_{ji}^{\delta} \right\| ^2}}
\nonumber	\\
&+\frac{5}{4}\beta \sum_{i=1}^{\mathcal{N}}{\sum_{j=1}^{\mathcal{N}}{a_{ij} \| C_ie_{ji}^{w} \| ^2}}
\end{align} 
In view of \eqref{e40} in Lemma \ref{lemma.3} of Section \ref{appendixB}, it gets
		\begin{align}\label{e23a}
		 \ell \dot{V}_2 \leqslant& \frac{1}{d_M}\sum_{i=1}^{\mathcal{N}}{\sum_{j=1}^{\mathcal{N}}{a_{ij}\dot{m}_{ij}\lambda _{M}^{-2}\left( C \right) \left( \left\| C_ie_{ji}^{\delta} \right\| ^2+\left\| C_je_{ji}^{w} \right\| ^2 \right)}}
			\nonumber\\
			&+\sum_{i=1}^{\mathcal{N}}{\sum_{j=1}^{\mathcal{N}}{a_{ij}m_{ij}}}\left( b_1\hat{c}_{ij}\left\| C_ie_{ji}^{\delta} \right\| ^2+b_2\left\| C_ie_{ji}^{w} \right\| ^2 \right) 
			\nonumber\\
			&+b_3 \tilde{\delta}^TC^T(\widehat{L}\otimes I)C\tilde{\delta}+\frac{1}{2}\rho \vartheta \lambda _{2}^{-1}\bar m \tilde{\delta}^TC^T({L}\otimes I)C\tilde{\delta}\nonumber\\
			&+\frac{1}{4}\beta \bar m \tilde{w}^TC^T(L\otimes I)C\tilde{w}
		\end{align}	
Accordingly, we can observe
$$\begin{aligned}
&(-\frac{1}{2}\alpha \phi+b_3)\tilde{\delta}^TC^T(\widehat{L}\otimes I)C\tilde{\delta}
\nonumber\\
&+\phi _{c} \sum_{i=1}^{\mathcal{N}}{\sum_{j=1}^{\mathcal{N}}{a_{ij}\left( c_{ij}-\bar c \right)}\| C_i\tilde{\delta}_i-C_j\tilde{\delta}_j\| ^2}
\nonumber\\
\leqslant&\phi _{c} \sum_{i=1}^{\mathcal{N}}{\sum_{j=1}^{\mathcal{N}}{a_{ij}} ({c}_{ij}-\hat {c}_{ij})\| C_i\tilde{\delta}_i-C_j\tilde{\delta}_j\| ^2}
\nonumber\\
&-\phi _{c} \bar c \sum_{i=1}^{\mathcal{N}}{\sum_{j=1}^{\mathcal{N}}{a_{ij}}\| C_i\tilde{\delta}_i-C_j\tilde{\delta}_j\|^2}
\nonumber\\
\leqslant&   -\phi _{c}(\bar c -\gamma_c ) \tilde{\delta}^TC^T(L\otimes I)C\tilde{\delta}
\end{aligned}$$ 	
Since $\hat c_{ij}\geqslant c_{ij}(0)$ and $c_{ij}(0) \geqslant 1$, it holds $\tilde{w}^TC^T(L\otimes I)C\tilde{w}\leqslant \tilde{w}^TC^T(\widehat{L}\otimes I)C\tilde{w}$. Combining \eqref{e23} and \eqref{e23a}, we have
		\begin{align}\label{e37}
\ell& \dot{V}\leqslant \mu _1\tilde{\delta}^TC^T(L\otimes I)C\tilde{\delta}-\frac{1}{4}\alpha \tilde{w}^TC^T(L\otimes I)C\tilde{w}
\nonumber\\
&+\sum_{i=1}^{\mathcal{N}}{\sum_{j=1}^{\mathcal{N}}{a_{ij}\left(\mu _2 \hat{c}_{ij}\left\| C_ie_{ji}^{\delta} \right\| ^2+\mu _3\left\| C_ie_{ji}^{w} \right\| ^2 \right)}}+\bar{\mathcal{H}}
		\end{align} 
		where $\mu _1=-\phi _{c}(\bar c -\gamma _c)+ 5\beta \bar{\phi}^2+\frac{1}{4}\alpha +\frac{1}{2}\rho \vartheta \lambda _{2}^{-1}\bar m+b_4$, $\mu _2=\hat{c}_{ij}^{-1}\lambda _{M}^{-2}\left( C \right) \dot{m}_{ij}+b_1m_{ij}+\frac{1}{2}\alpha \phi$, $\mu _3=\lambda _{M}^{-2}\left( C \right) \dot{m}_{ij}+b_2m_{ij}+\frac{5}{4}\beta$.
		
The two cases are given to analysis inequality \eqref{e37}:
		
		\emph{Case 1}: The time interval when activation function is active, i.e., $\dot{m}_{ij}\ne 0$, $t\in [t_{k}^{ij},t_{k}^{ij}+t_{\rm MEI}^{ij})$. 
		Note that $\mu _2 \leqslant 0$ and $\mu _3\leqslant 0$. Moreover, since $\bar c$ can be any large value, there exists a constant $\bar c$ such that $\mu _1\leqslant-\frac{1}{4}\alpha $. 
		For this reason, we have
		\begin{align}\label{e37b}
			\ell \dot{V}\leqslant& -\frac{1}{4}\beta \tilde{\delta}^TC^T(L\otimes I)C\tilde{\delta}-\frac{1}{4}\beta\tilde{w}^TC^T(L\otimes I)C\tilde{w}+\bar{\mathcal{H}}
		\end{align} 
		
		\emph{Case 2}: The time interval when activation function is dormant, i.e., $\dot m_{ij}=0$, $t\in [t_{k}^{ij}+t_{\rm MEI}^{ij},t_{k+1}^{ij})$.
		Applying the ETC in \eqref{e4}, one gets
		\begin{align}\label{e39a}
\sum_{i=1}^{\mathcal{N}}{\sum_{j=1}^{\mathcal{N}}{a_{ij}\hat c_{ij}\| C_ie_{ji}^{\delta}\| ^2}}\leqslant \sum_{i=1}^{\mathcal{N}}{\sum_{j=1}^{\mathcal{N}}{a_{ij}\hat c_{ij}\gamma_\delta }\leqslant}d_M\mathcal{N} \gamma_\delta, 
		\end{align} and
		\begin{align}\label{e38}
			\sum_{i=1}^{\mathcal{N}}{\sum_{j=1}^{\mathcal{N}}{a_{ij}\left\| C_ie_{ji}^{w} \right\| ^2}}&\leqslant \sum_{i=1}^{\mathcal{N}}{\sum_{j=1}^{\mathcal{N}}{a_{ij}\gamma_w}} \leqslant d_M\mathcal{N} \gamma_w
		\end{align} 
	By \eqref{e39a} and \eqref{e38}, the inequality \eqref{e37} can be derived as
		\begin{align}\label{e25}
			\ell \dot{V}\leqslant& \mu _1\tilde{\delta}^TC^T(L\otimes I)C\tilde{\delta}+\mu _2\tilde{w}^TC^T(L\otimes I)C\tilde{w}
			\nonumber\\
			&+\mu _4 \gamma_\delta +\mu _5 \gamma_w +\bar{\mathcal{H}}
		\end{align} 
		with $\mu _4=\left( \frac{1}{2}\alpha \phi +b_1 \bar m \right) d_M\mathcal{N}$, $\mu _5=\left( \frac{5}{4} \beta +b_2 \bar m \right) d_M\mathcal{N}$, where we utilize the condition $m_{ij}\leqslant \frac{1}{\hat{c}_{ij}}m_{ji} (0)\leqslant \bar{m}$.
		Similarly, it is not difficult to derive that $\mu _1\leqslant -\frac{1}{4}\beta $. Then \eqref{e25} can be rewritten as
		\begin{align}\label{e37c}	
			\ell \dot{V}\leqslant& -\frac{1}{4}\beta \tilde{w}^TC^T(L\otimes I)C\tilde{w}-\frac{1}{4}\beta \tilde{\delta}^TC^T(L\otimes I)C\tilde{\delta}
			\nonumber\\
			&+\mu _4 \gamma_\delta +\mu _5 \gamma_w +\bar{\mathcal{H}}
		\end{align} 
	With the property of set-valued Lie derivative, we can obtain from \eqref{e37b} and \eqref{e37c} that $\forall t\in [t_{\varepsilon}^r, t_{\varepsilon+1}^r)$,
		\begin{align}\label{e31}
			V\left( t \right) \leqslant& V\left( t_{\varepsilon}^{r} \right)-\frac{1}{4}\beta \lambda _2\lambda _m^2 (C) \int_{t_{\varepsilon}^{r}}^t{\left( \|\tilde{w}\left( \tau \right)\|^2+ \|\tilde{\delta}\left( \tau \right) \| ^2d\right)\tau}
			\nonumber\\
			&+\int_{t_{\varepsilon}^{r}}^t{\left(\mu _4\gamma_\delta \left( \tau \right)+ \mu _5\gamma_w \left( \tau \right) +\bar{\mathcal{H}}\left( \tau \right)\right) d\tau}
		\end{align}
		Since at  time instant $t_{\varepsilon+1}^{r}$, at least one Byzantine agent is isolated.
		It results in the boundedness of $\int_{t_{\varepsilon}^{r}}^{t_{\varepsilon+1}^{r}}{ \bar{\mathcal{H}} \left( \tau \right)}d\tau$. By \eqref{e15a} and \eqref{e15b}, it follows that both $\digamma _{i}^{\delta}$ and $\digamma _{i}^{w}$ continue to decrease to zero as $t \rightarrow \infty$. Since the tampered values of Byzantine agents will not be zero, it allows each agent to ultimately identify any Byzantine agents even if their tampered data (i.e., $\hat \delta_i$ and $\hat w_i$) is slightly deviated from the real value. Since the number of Byzantine agents is finite, there exists an instant $t_\hbar^r$ when all Byzantine agents are isolated by all their neighbors. Since $\mathcal{G}$ is $(r+1)$-connected, from Lemma \ref{lemma.2}, one gets that $\mathcal{G}$ is $r$-isolatable, which indicates that $\mathcal{G_N}$ is connected. Then, $\forall t\in [t_{\hbar}^{r}, \infty)$,
	\begin{align}\label{e32}
		V\left( t \right) \leqslant& -\frac{1}{4}\beta \lambda _2\lambda _m^2(C)\int_{t_{\hbar}^{r}}^t{\left(\| \tilde{w}\left( \tau \right)\|^2+\| \tilde{\delta}\left( \tau \right)\| ^2\right)d\tau}
\nonumber\\
&+\int_{t_{\hbar}^{r}}^t{\left(\mu _4\gamma_\delta \left( \tau \right)+ \mu _5\gamma_w \left( \tau \right)\right) d\tau}+V\left( t_{\hbar}^{r} \right)
		\end{align}
	Since $\gamma_\delta, \gamma_w \in \mathcal{L} _1$, by \eqref{e31} and \eqref{e32}, it results in that $V(t)$ is bounded for $t\in [0, \infty)$. From Lemma \ref{lemma.6}, we have that $\tilde{\delta}$ and $\tilde{w}$ will convergent to zero as $t \rightarrow \infty$. Obviously, $\lim_{t\rightarrow \infty} \delta_i=\delta_i^*$, $\lim_{t\rightarrow \infty} w_i=w_i^*$, and each adaptive coupling weight $c_{ij}$ will finally convergent to a steady value. Considering equation \eqref{e16a}, it follows that there exists a positive constant $T_{\mathrm{MEI}}^{ji}<T_0$, indicating that Zeno behavior is effectively excluded. $\hfill\blacksquare$

\begin{remark}\label{re02}
	Note that the value $T_0$ depending on the coupling weight $\hat c_{ij}$ will be changed at each trigger instant. To avoid its continuous updating, we give a fixed $\hat T_0$ in a time interval. One can define a piecewise function $\kappa $ satisfying $\kappa =c_{ij}\left( 0 \right) + v\bar{q}$ for $\hat{c}_{ij}\in \left[ c_{ij}\left( 0 \right) + v\left( \bar{q}-1 \right), c_{ij}\left( 0 \right) +v \bar{q} \right)$, where $v>0$, $\bar{q}\in \mathbb{Z} ^+$. Obviously, it always holds that $\kappa >\hat{c}_{ij}$. Then we introduce an auxiliary function  $\dot{\hat{m}}_{ji}=-\hat{s}_{ji}(\bar{\sigma}_1\hat{m}_{ij}+\bar{\sigma}_2)$ where $\bar{\sigma}_1=\lambda _{M}^{2}\left( C \right) \bar{b}\kappa ,\bar{\sigma}_2=\frac{1}{2}\lambda _{M}^{2}\left( C \right) \alpha \phi \kappa$, $\hat{s}_{ji}$ is the same as that in \eqref{e16}, $\hat{m}_{ji}(t_{k}^{ji+})=\frac{1}{\hat{c}_{ij}}\hat m_{ji}(0),\hat{m}_{ji}(0)=m_{ji}(0)$. Consequently, there exists a positive $\hat T_0$ satisfying
	\begin{align}
	\hat{T}_0=&\frac{1}{\lambda _{M}^{2}\left( C \right) b_M\kappa}\ln \left( \frac{1}{2}+\frac{1}{2}\sqrt{1+\frac{8b_M}{\alpha \phi \kappa}m_{ji}(0)} \right) \nonumber
	\end{align}
for $\hat{c}_{ij}\in \left[ c_{ij}\left( 0 \right) + v\left( \bar{q}-1 \right), c_{ij}\left( 0 \right) +v \bar{q} \right)$. In view of \eqref{e16}, it is not difficult to obtain that $\hat{T}_0<T_0$. As thus, one can choose a fixed value $T_{\mathrm{MEI}}^{ji}$ to guarantee $T_{\mathrm{MEI}}^{ji} \leqslant \hat{T}_0$, for $\hat{c}_{ij}\in \left[ c_{ij}\left( 0 \right) + v\left( \bar{q}-1 \right), c_{ij}\left( 0 \right) +v \bar{q} \right)$.
Furthermore, we can set a small $\eta_{ij}$ to mitigate the growth of $c_{ij}$, which in turn enables a  bigger
$\hat{T}_0$.
\end{remark} 

\subsection{Convergence of Output Trajectory Tracking}
In light of \eqref{e3}, the structure of the distributed controller can be reformulated as
\begin{align}\label{e5}
	u_i=&-\bar{B}_i\left[ F_i\bar{K}_i\left( \delta _i-x_i \right) +\bar{K}_i( \dot \delta _i-A_ix_i) \right] 
	\nonumber\\
	=&\bar{B}_iF_i\bar{K}_i\left( x_i-\delta _i \right) +\bar{B}_i\bar{K}_i ( A_ix_i-\dot \delta _i )
\end{align}   
Let	$x_{i}^{*}=\mathrm{col}\{ x_{i,1}^{*},x_{i,2}^{*} \} $ be a equilibrium point of the HMAS in \eqref{e1}. After that, we denote
$\chi_{i,1}=x_{i,1}-x_{i,1}^{*}$, $\chi_{i,2}=-K_i\left( x_{i,1}-\delta _{i,1} \right) +\left( x_{i,2}-\delta _{i,2} \right)=\bar{K}_i\left( x_i-\delta _i \right)$. Taking the derivative of $\chi_{i,1}$ and $\chi_{i,2}$ yields
$$\begin{aligned}
	\dot{\chi}_{i,1}=&A_{i,11}\chi _{i,1}+A_{i,12}\chi _{i,1}
	\\
	\dot{\chi}_{i,2}=&\bar{K}_i\left( A_ix_i+B_iu_i-\varphi _i \right) 
	\\
	=&\bar{K}_iA_ix_i+F_i\bar{K}_i\delta _i-\left( \bar{K}_iA_i+F_i\bar{K}_i \right) x_i
	\\
	=&F_i\bar{K}_i\left( \delta _i-x_i \right)
\end{aligned}
$$	
\begin{theorem}\label{theoreom.2}
	For HMAS \eqref{e1}, Problem $\ref{pro1}$ can be solved, if $\lim _{t\rightarrow \infty}\delta _i=\delta _{i}^{*}$, $\lim _{t\rightarrow \infty}w_i=w_{i}^{*}$ and there exists a symmetric matrix $P_i>0$, $i \in \mathcal{I}$, such that
	\begin{align}
		\varLambda _{i}^{T}P_i+P_i\varLambda _i+2P_i^2-\bar \mu I< 0
	\end{align}
	with $\bar \mu >0$, and $\varLambda _i\triangleq A_{i,12}K_i$ being Hurwitz.
\end{theorem}

\emph{Proof}: See Section \ref{appendixC}. $\hfill\blacksquare$
\begin{remark}
Different from the related works on distributed optimization in \cite{li2022exponential,xian2024distributed, zhang2022optimal} that depend on the eigenvalues of the Laplacian matrix, the designed adaptive optimal protocol \eqref{e3} does not need such information and the parameters $\rho$, $\alpha$ and $\beta$ can be arbitrary positive values. This feature is beneficial for solving optimization problems under the unknown change of communication topology.  Although the existing works in \cite{wu2018distributed,li2019distributed,xian2023robust} are independent of the eigenvalue of the Laplacian matrix, they cannot be applied to the case involving unknown changes of communication topology, as their protocols are a type of proportional integral protocols. Furthermore, the optimal protocols for HMASs in \cite{li2022exponential, xian2024distributed, zhang2022optimal}, \cite{xian2023robust} are limited to a class of full-actuated heterogeneous system dynamics, in contrast to the more general systems studied in this paper.
\end{remark}
%\begin{remark}
%	In contrasted to the MSR-based distributed resilient optimization protocols in \cite{su2020byzantine,wu2023byzantine,kuwaranancharoen2020byzantine}, it is not easy for normal agents to reach precise optimization results under the influence of Byzantine attackers. Note that the MSR-based protocols rely on ($2f+1$)-robustness graph, and all normal agents can merely reach to a sub-optimal point within a convex hull that encompasses the optimal point.
%\end{remark}
\begin{remark}
In contrast to the event-triggered optimization protocols of MASs presented in \cite{li2022exponential},\cite{wu2018distributed,li2019distributed,xian2023robust,dai2020distributed}, which feature triggering intervals that tend toward zero as the agents' states approach their equilibrium, the designed event-triggered protocol has the time-varying, strictly positive MEI that can be calculated by agents at the trigger instant. The presence of MEI significantly reduces the communication frequency required between agents. Notably, to the best of our knowledge, this paper is the first to explore the event-triggered distributed optimization problem for MASs incorporating the MEI.
\end{remark}

\begin{remark}
	%It should be noted that since the use of adaptive coupling weight $c_{ji}$, the gain parameters $\alpha, \beta, \rho$ affecting the iteration rates of $\delta_i$ and $w_i$ can be any positive values. 
	When some Byzantine agents are isolated by other normal agents, the communication topology will be changed. Under such circumstances, the transient change rates of $\delta_i$ and $w_i$ will be quite large, leading to the requirement of high iterative accuracy of $\delta_i$ and $w_i$ in a short time; otherwise, it is prone to false detection when the thresholds $\digamma _{ij}^{\delta}$ and $\digamma _{ij}^{w}$ become small, potentially leading to the isolation of normal agents. As a result, by choosing the small gain parameters $\rho$, $\alpha$, $\beta$, the requirement for high iterative accuracy of $\delta_i$ and $w_i$ can be mitigated. However, those gain parameters in works \cite{li2022exponential},\cite{li2019distributed,xian2023robust,dai2020distributed} can not be arbitrarily small because of they require the eigenvalues of the Laplacian matrix.
\end{remark}

\section{Numerical Example}	
	In this section, to illustrate the validity of the proposed detection based distributed resilient protocol for optimization, we examine a scenario involving eight heterogeneous mobile robots tasked with an optimal assembly problem. 

%\subsection{Robot dynamics and optimization objective}	
	Initially, the communication topology is depicted as shown in Fig. \ref{fig1} (b), wherein agents 
	$1$ and $2$ are identified as the Byzantine robots capable of distorting their true information when communicating with their neighbors. The system dynamic of each mobile robot in \eqref{e1}, as described by \cite{an2021distributed}, is given by
\begin{align}
	\begin{bmatrix}
		\dot{\mathbf{p}}_i \\
		\dot{\mathbf{v}}_i \\
	\end{bmatrix} 
	=& 
	\begin{bmatrix}
		0 & I_2 \\
		0 & -\frac{\mathcal{T}_i}{\mathcal{W}_i} \\
	\end{bmatrix}
	\begin{bmatrix}
		\mathbf{p}_i \\
		\mathbf{v}_i \\
	\end{bmatrix}
	+ 
	\begin{bmatrix}
		0 \\
		\frac{\mathcal{T}_i}{\mathcal{W}_i} \\
	\end{bmatrix} u_i\nonumber\\
\mathbf{y}_i=&\left[ \begin{matrix}
	I_2&		0\\
\end{matrix} \right] \begin{bmatrix}
	\mathbf{p}_i\\
	\mathbf{v}_i\\
\end{bmatrix} \nonumber
\end{align}
where $\mathbf{p}_i = {\rm col}\{ p_{i}^{x}, p_{i}^{y} \} \in \mathbb{R}^2$ and $\mathbf{v}_i = {\rm col}\{ v_{i}^{x}, v_{i}^{y} \} \in \mathbb{R}^2$. For agent $i$, $p_{i}^{x}$ and $p_{i}^{y}$ denote the lateral and longitudinal positions, respectively, while $v_{i}^{x}$ and $v_{i}^{y}$ represent the lateral and longitudinal velocities. Additionally, $\mathcal{T}_i$ signifies the friction coefficient, and $\mathcal{W}_i$ denotes the mass of the robot.
\begin{table}[t]
	\centering
	\caption{Parameter of agents}
	\label{table:positions}
	\setlength{\tabcolsep}{3pt} % 调整列间距
	\renewcommand{\arraystretch}{1} % 调整行间距（可选）
	\resizebox{8cm}{!}{%
		\begin{tabular}{|c|>{$}l<{$}|>{$}l<{$}|>{$}l<{$}|}
			\hline
			Agent & \mathrm{col}\{\mathbf{p}_i(0), \mathbf{v}_i(0)\} & \delta_i(0) & w_i(0) \\
			\hline
			1 & \mathrm{col}\left\{ 1.0,-0.5,0,0 \right\} & \mathrm{col}\left\{1.0,-0.5,0,0  \right\} & \mathrm{col}\left\{ -0.2, -0.1, 0, 0.2 \right\} \\
			2 & \mathrm{col}\left\{ 0.5, 1.0,0,0 \right\} & \mathrm{col}\left\{0.5, 1.0,0,0 \right\} & \mathrm{col}\left\{  -0.1, 0.2, 0.1, 0.3  \right\} \\
			3 & \mathrm{col}\left\{ 1.5, -1.0,0,0 \right\} & \mathrm{col}\left\{  1.5, -1.0,0,0 \right\} & \mathrm{col}\left\{ 0.1, 0.2, -0.4, 0.1 \right\} \\
			4 & \mathrm{col}\left\{ -0.5, 0.5,0,0 \right\} & \mathrm{col}\left\{ -0.5, 0.5,0,0 \right\} & \mathrm{col}\left\{  0.4, -0.2, -0.3, 0.1 \right\} \\
			5 & \mathrm{col}\left\{  0.5, -1.0,0,0 \right\} & \mathrm{col}\left\{ 0.5, -1.0,0,0  \right\} & \mathrm{col}\left\{-0.4, 0, 0.2, 0.1 \right\} \\
			6 & \mathrm{col}\left\{ 1.0, 1.0,0,0 \right\} & \mathrm{col}\left\{1.0, 1.0,0,0 \right\} & \mathrm{col}\left\{0.2, 0.1, 0, 0.3\right\} \\
			7 & \mathrm{col}\left\{ -1.0, 1.5,0,0 \right\} & \mathrm{col}\left\{-1.0, 1.5,0,0 \right\} & \mathrm{col}\left\{ 0.1, -0.2, -0.2, 0.3  \right\} \\
			8 & \mathrm{col}\left\{ 0.5, 0.5,0,0 \right\} & \mathrm{col}\left\{0.5, 0.5,0,0  \right\} & \mathrm{col}\left\{0.2, 0.3, -0.1, 0.1 \right\} \\
			\hline
		\end{tabular}%
	}
\end{table}
\begin{figure}[!t]
	\centering
	\vspace{0.2cm}
	\subfigtopskip=1pt 
	\subfigbottomskip=1pt
	\subfigcapskip=2pt
	\includegraphics[width=0.42 \textwidth]{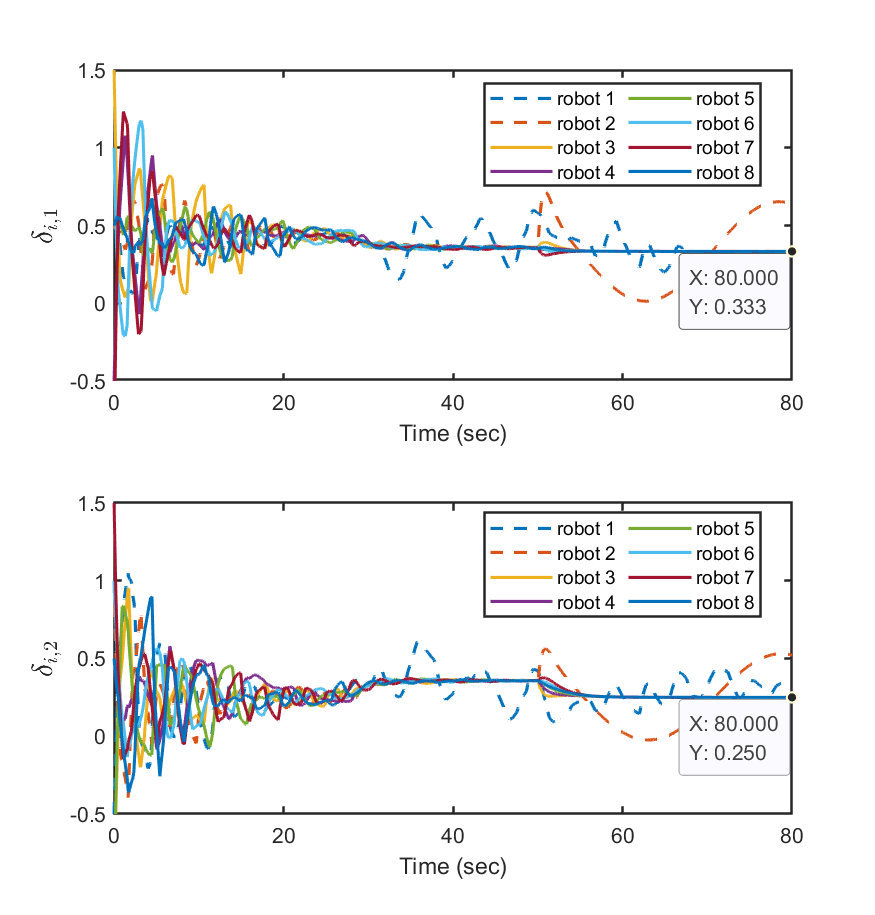}
	\caption{ Curves of $\delta_i$ in 0-80 s. }\label{fig4} 
\end{figure}
\begin{figure}[!t]
	\centering
	\vspace{0.2cm}
	\subfigtopskip=1pt 
	\subfigbottomskip=1pt
	\subfigcapskip=2pt
	\includegraphics[width=0.42 \textwidth]{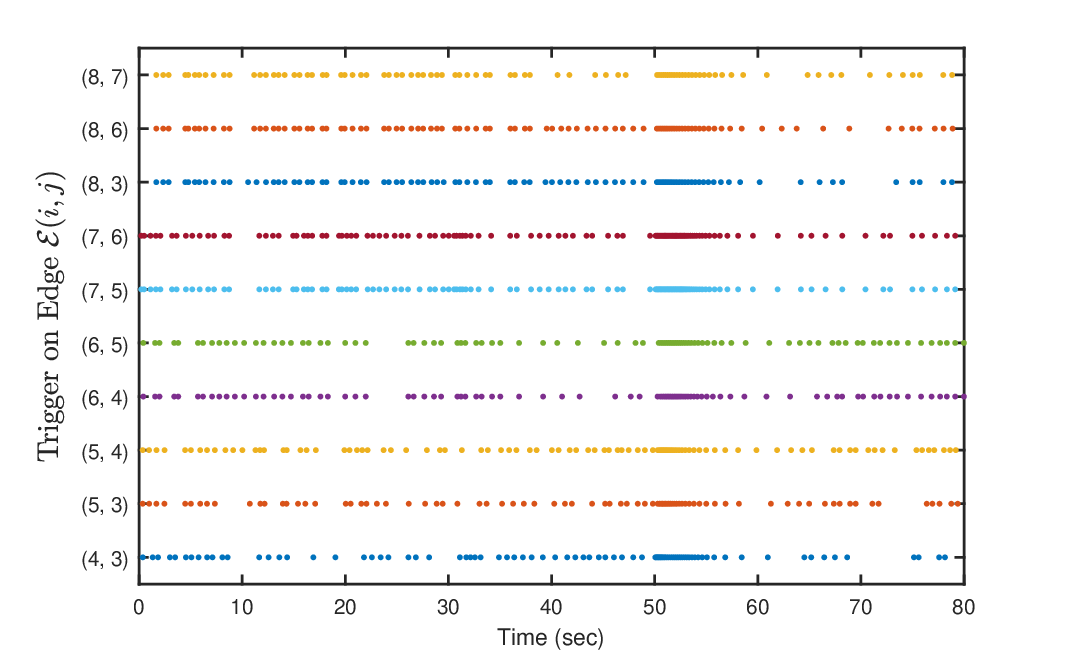}
	\caption{ Triggering instants of the edges among normal agents. }\label{fig5} 
\end{figure}

Under Assumption \ref{ass3}, the optimization problem of all eight robots is given by
\begin{align}
&\underset{y_i\in \mathbb{R}^{q}}{\min}\sum\nolimits_{i \in \mathcal{I}_\mathcal{N}}^{}{\left\| y_i(t)-y_i\left( 0 \right) \right\| ^2,} \nonumber \\
&\mathrm{s. t.}\,\, y_i=y_j,\ [A_{i,11}\,\,A_{i,12}]y_i=0\nonumber
\end{align}
where $ \forall i,j\in \mathcal{I_N}$. Therefore, the optimization  solution of all the normal robots is $\mathrm{col}\{0.333, 0.250\}$.
For agent $i$, choose $\mathcal{T}_i=I_2 \otimes (1-0.1i)$, $\mathcal{W}_i= 0.2\sin(2i)$. The initial values of other parameters are listed in Table I. Then select $\rho =0.1$, $\alpha =0.5$, $\beta=0.5$, $F_i=1.5I_2$, $K_i=-I_2$, $a_{ij}=1$, $c_{ij}=1$, $\eta_{ij}=0.02$, $\gamma _{\delta}=\gamma _w=e^{-0.2t}$, $\gamma _{c}=0.1$. Consequently, we have $d_M=5$. Following the method of Remark \ref{re02},  we design $\hat m_{ij}(0)=1$, $\bar{\sigma}_1=3\kappa ,\bar{\sigma}_2=\frac{5}{8}\kappa$, ${v}=0.2$, under which one has that if $\hat c_{ij}<1.2$, it holds $\kappa=1.2$. Therefore, $t_{\rm MEI}^{ij}$ can be set to $0.10 s$ for $\hat c_{ij}<1.2$. In addition,we select the time-varying thresholds $\digamma _{ij}^{\delta}=\digamma _{ij}^{w}=10^{-3}+1.2e^{-0.15t}$.

\begin{figure}[!t]
	\centering
	\vspace{0.2cm}
	\subfigtopskip=1pt 
	\subfigbottomskip=1pt
	\subfigcapskip=2pt
	\includegraphics[width=0.38 \textwidth]{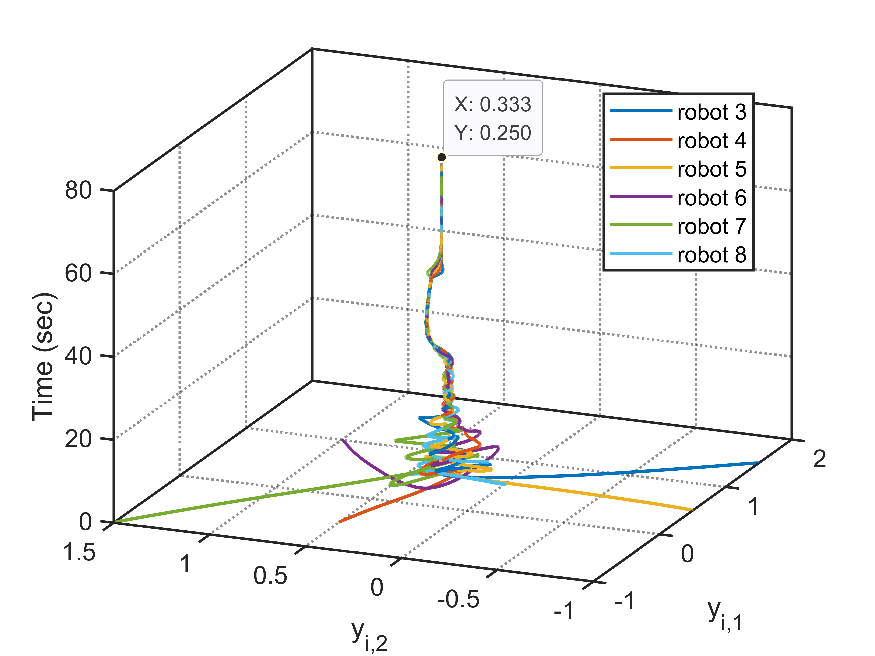}
	\caption{Curves of the output states of all normal robots.}\label{fig6} 
\end{figure}

\begin{figure}[!t]
	\centering
	\vspace{0.2cm}
	\subfigtopskip=1pt 
	\subfigbottomskip=1pt
	\subfigcapskip=2pt
	\includegraphics[width=0.35 \textwidth]{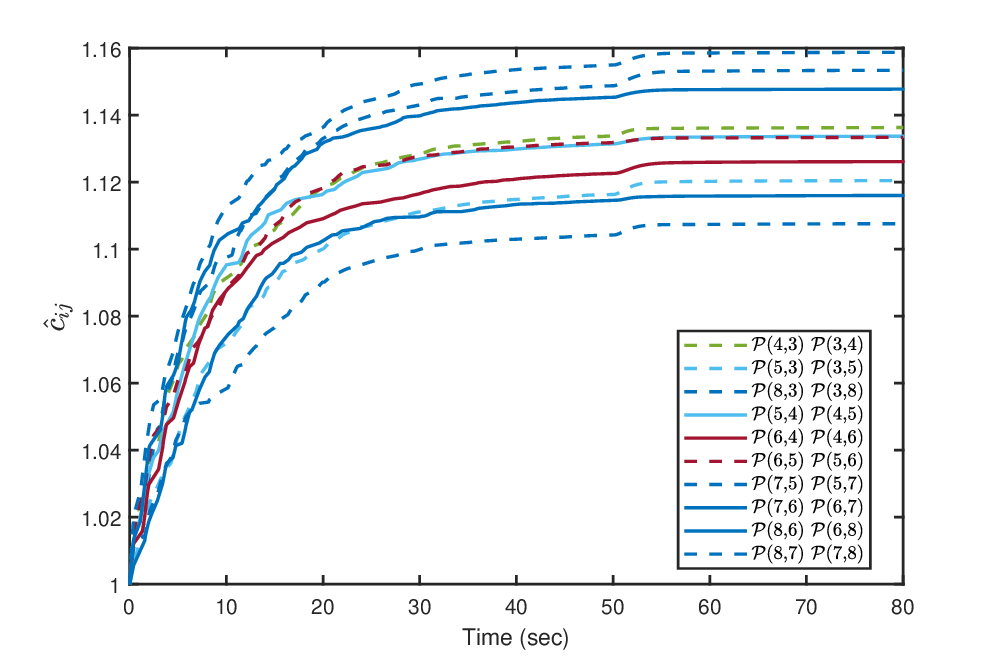}
	\caption{Trajectories of coupling weights $\hat c_{ij}$ for all normal robots.}\label{fig8} 
\end{figure}
We now illustrate the detailed information for Byzantine agents, i.e., robots 1 and 2. First, assume that robot $1$,  the Byzantine attacker, is unaware of its own MEI, which can randomly trigger at any time and may occur within the MEI. Furthermore, it can also tamper with the true values $C_1 \hat \delta_{j1}$, $C_1 \hat w_{j1}$ with the corresponding deviations given by $C_1\epsilon _{j1}^{\delta}=0.002j\left[ \begin{smallmatrix}
	\sin \left( 0.02jt \right)\\
	\sin \left( 0.02jt \right)\\
\end{smallmatrix} \right]$, $C_1\epsilon _{j1}^{w}=0.002j\left[ \begin{smallmatrix}
	\cos \left( 0.02jt \right)\\
	\cos \left( 0.02jt \right)\\
\end{smallmatrix} \right]$, $t\in \left[20,30 \right) s;$ and $C_1\epsilon _{j1}^{\delta}=0.02j\left[ \begin{smallmatrix}
\sin \left( 0.2jt \right)\\
\sin \left( 0.2jt \right)\\
\end{smallmatrix} \right]$, $C_1\epsilon _{j1}^{w}=0.1j\left[ \begin{smallmatrix}
\cos \left( 0.2jt \right)\\
\cos \left( 0.2jt \right)\\
\end{smallmatrix} \right]$, $t\in \left[ 30,80 \right) s$, where $j\in \mathcal{N}_1$. Here the set of minor deviations $C_1\epsilon _{j1}^{\delta}$ and $C_1 \hat w_{j1}$ in $\left[20,30 \right) s$ is aimed at avoiding the detection of TEC.
As for robot $2$, it behaves cautious so that it will not trigger within the MEI, and only temples with the data in its own trigger times, and the deviations of $C_2 \hat \delta_{j2}$, $C_1 \hat w_{j2}$, $j\in \mathcal{N}_2$, are designed by
$	C_2\epsilon_{j2}^{\delta}= 0.05j \left[\begin{smallmatrix}
		\cos(0.2t) \\
		\cos(0.2t)
	\end{smallmatrix}\right]$, $t \in [30,80) s$,
	$C_2\epsilon_{j2}^{w}= 0.2\sqrt{j} \left[\begin{smallmatrix}
		\sin(0.2jt) \\
		\sin(0.2jt)
	\end{smallmatrix}\right]$,  $t \in [30,80) s$, for $j\in \mathcal{N}_2$.
 %Byzantine robots 1 and 2 can tamper with their values and transmit the altered information within the TEI

\begin{figure}[!t]
	\centering
	\vspace{0.2cm}
	\subfigtopskip=1pt 
	\subfigbottomskip=1pt
	\subfigcapskip=2pt
	\subfigure[]{\includegraphics[width=0.4 \textwidth]{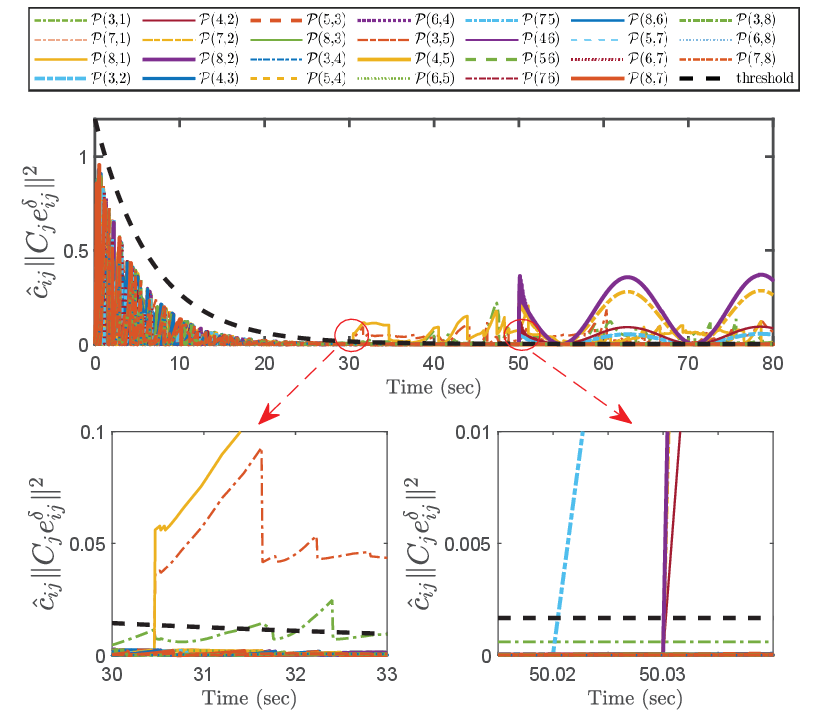}}
	\subfigure[]{\includegraphics[width=0.4 \textwidth]{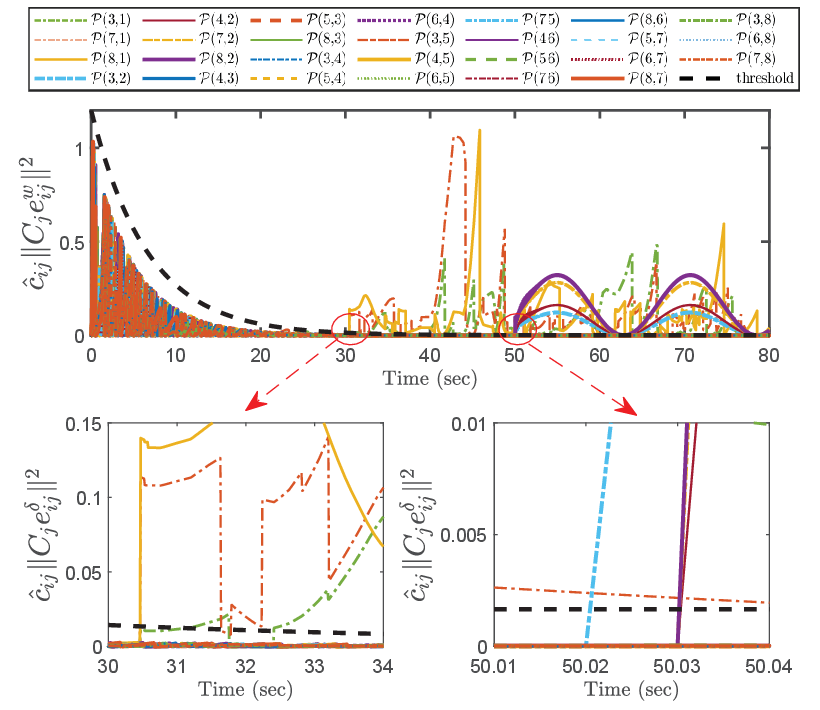}}
	\caption{ Curves of triggering errors and thresholds on path $\mathcal{P} \left( i,j \right)$, where $i \in \mathcal{I_N}$, $j \in \mathcal{N}_i \cup \{1, 2\}$. (a) Curves of $\hat c_{ij}\| C_je_{{ij}}^{\delta}\| ^2$ and $\digamma _{ij}^{\delta}$. (b) Curves of $\| C_je_{{ij}}^{w}\| ^2$ and $\digamma _{ij}^{w}$.}\label{fig7} 
\end{figure}

\begin{figure}[!t]
	\centering
	\vspace{0.2cm}
	\subfigtopskip=1pt 
	\subfigbottomskip=1pt
	\subfigcapskip=2pt
	\includegraphics[width=0.42 \textwidth]{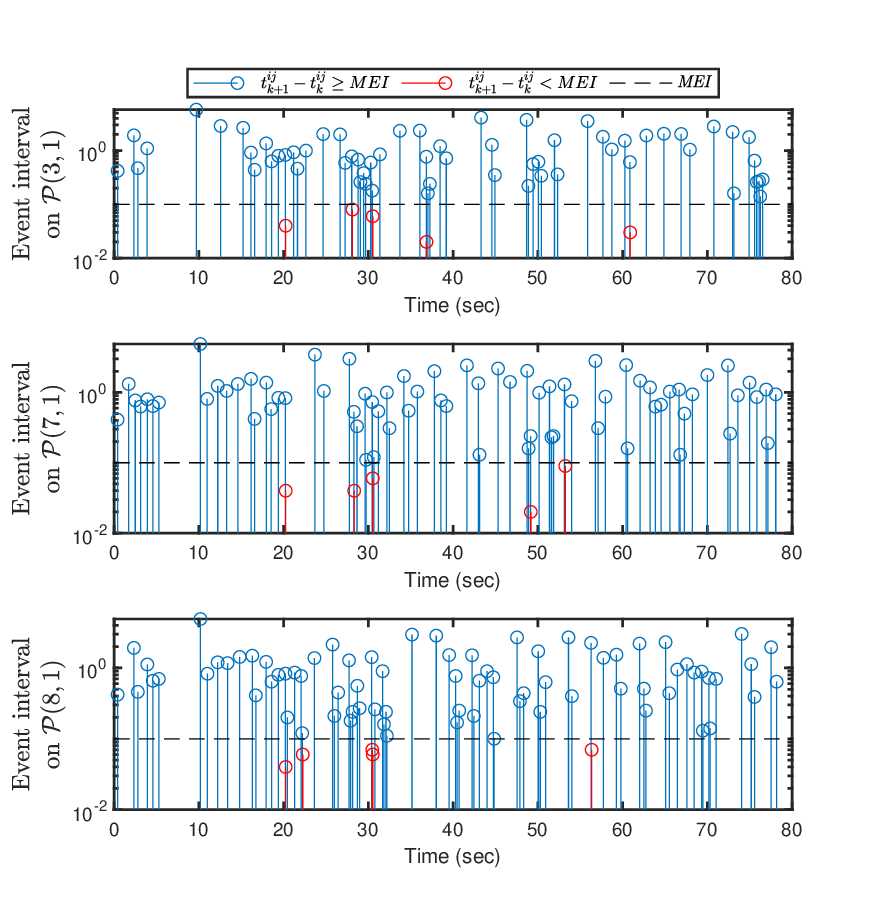}
	\caption{Trigger interval information on $\mathcal{P}(3,1)$,  $\mathcal{P}(7,1)$,  $\mathcal{P}(8,1)$. Note that agent 1 is the Byzantine agent and agents 3,7,8 are normal agents.}\label{fig9} 
\end{figure}

In this experiment, we trial the proposed optimization protocol in $[0,\ 80) s$, and the discretization time of the MASs is $10^{-4} s$. On this occasion, agent $1$ starts transmitting tampered values after $20 s$, while agent $2$ transmits tampered values after $50 s$. In Fig. \ref{fig4}, we give the curves of the optimal observer $\delta_i$ form $0 s$ to $80 s$. Under the designed BADI algorithm, it is obvious that robots $1$ and $2$ can be successfully isolated after $20 s$  and $50 s$, respectively, and finally, all $\delta_i$ of the normal agents can estimate their desired optimal values. The position trajectories of the six normal robots in $0$-$80 s$ are depicted in Fig. \ref{fig6}, from which one can verify all the normal robots converge to their optimal positions. Additionally, the trajectories of  all the coupling weights $\hat c_{ij}$ for normal robots are plotted in Fig. \ref{fig8}. As shown in Fig. \ref{fig7}, robot $i$ can detect if the transmitted parameters are available from robot $j$ via path $\mathcal{P}(i,j)$. In addition, we present the triggering information between agent $1$ and its neighbors in Fig. \ref{fig9}, which depicts that after $20 s$, several successive triggering intervals satisfying $t_{k+1}^{ij}-t_{k}^{ij}<t_{\rm MEI}^{ij}$ will be detected under the use of BADI Algorithm. By Figs. \ref{fig7} and \ref{fig9}, it is evident that all the neighbors of robot $1$ initially detect the Byzantine robot $1$ after $20 s$ under TIC, which is quicker than that under TEC.
%
%\textit{Comparison}: Here, we consider the scenario where the detection mechanism includes only the error bounds condition, excluding the time trigger interval condition. All other settings remain the same as in the previous case. Fig. \ref{fig11} gives the curves of of triggering error, where we can see that under the BAD Algorithm without the trigger interval condition, the detector is not so sensitive such that it can will not detect the Byzantine robots in $[15, 20) s$.

%\begin{figure}[!t]
%	\centering
%	\vspace{0.2cm}
%	\subfigtopskip=1pt 
%	\subfigbottomskip=1pt
%	\subfigcapskip=2pt
%	\includegraphics[width=0.42 \textwidth]{fig11.eps}
%	\caption{Error Curves of two successive triggers for all normal robots.}\label{fig11} 
%\end{figure}

\section{Conclusion}
This paper addressed the distributed resilient optimization problem of HMAS under adversarial agents. The edge-based adaptive event-triggered optimization protocol has been proposed, which does not require knowledge of the nonzero smallest eigenvalue of the Laplacian matrix. With such event-triggered mechanism, the agent on each communication edge has the known time-varying positive MEI. A novel self-triggered hybrid detection approach has been designed, which combines both the triggering interval condition and error-bounded condition. This approach enables agents to identify potential Byzantine agents among their neighbors and subsequently sever the corresponding communication links. Future research interests will address on distributed resilient optimization under attacks over unbalanced directed graphs.

\section{Appendix}
\subsection{Proof of Lemma \ref{lemma.2}} \label{appendixA}
Let set $\mathcal{S} _1=\left\{ j_1 \right\}, \forall j_1 \in \mathcal{V} $. Since $\mathcal{G}$ is $(r+1)$-connected, agent $j_1 \in \mathcal{S}_1$ has at least $r+1$ neighbors in $\mathcal{V} \setminus \mathcal{S}_1$. If the $r$ neighbors of agent $j_1$ are removed, there exists at least one neighbor for agent $j_1$. This directly implies that there must exist a bidirectional path between $j_2 \in \mathcal{V} \setminus \mathcal{S}_1$ and agent $j_1$.

Define $\mathcal{S} _2=\mathcal{S} _1\cup \left\{ j_2 \right\}$. Because $\mathcal{G}$ is $(r+1)$-connected, one can get that there exists at least one agent in $\mathcal{S} _2$ having $r+1$ neighbors. Since the agent in $\mathcal{S} _1$ (i.e. agent $j_1$) can be any node in $\mathcal{V}$, it can not guarantee the existence of $r+1$ neighbors in $\mathcal{V} \setminus \mathcal{S} _2$. Hence, one gets that $j_2$ has at least $r+1$ neighbors in $\mathcal{V} \setminus \mathcal{S} _2$.  Therefore, if isolating $r$ neighbors in $\mathcal{V} \setminus \mathcal{S} _2$ for agent $j_2$, there must exist one neighbor $j_3$ in $\mathcal{V} \setminus \mathcal{S} _2$. It thus leads to that there exists a bidirectional path between agents $j_3$ and $j_1$.

Define $\mathcal{S} _{\iota}=\mathcal{S} _1\cup \left\{ j_2 \right\} ...\cup \left\{ j_{\iota} \right\} ,\iota =2,...,N$. Taking a similar step, one can verify that after isolating $r$ agents for agent $j_{\iota}$ in $\mathcal{S} _{\iota}$,  $j_{\iota}$ has at least one neighbor $j_{\iota+1}$ in $\mathcal{V} \setminus \mathcal{S} _{\iota}$. When the cardinality of $\mathcal{V} \setminus \mathcal{S} _{\iota}$ becomes $r+1$, one has that there exists at least one agent after isolating $r$ agents, and that agent is a neighbor of agent $j_{\iota}$.  Indeed, it follows that there is a bidirectional path between agents $j_{N}$ and $j_{1}$, which thus indicates that the graph is $r$-isolatable.

\subsection{Lemma \ref{lemma.3}}\label{appendixB}
\begin{lemma}\label{lemma.3}
For the designed Lyapunov function $V_2$ in \eqref{e25th_b}, its set-valued Lie derivative satisfies that 
	\begin{align}
		\ell& \dot{V}_2\leqslant \frac{1}{d_M} \sum_{i=1}^{\mathcal{N}}{\sum_{j=1}^{\mathcal{N}}{a_{ij}\dot{m}_{ij}\lambda _{M}^{-2}\left( C \right) \left( \left\| C_ie_{ji}^{\delta} \right\| ^2+\left\| C_je_{ji}^{w} \right\| ^2 \right)}}
		\nonumber\\
		&+\Big(3\alpha+\frac{5}{2}\beta + \frac{2\bar \eta \gamma_{c}}{d_M}+\frac{\rho }{2d_M}\Big)\sum_{i=1}^{\mathcal{N}}{\sum_{j=1}^{\mathcal{N}}{a_{ij}\hat{c}_{ij}m_{ij}}}\left\| C_ie_{ji}^{\delta} \right\| ^2
		\nonumber\\
		&+(\alpha+\beta) \sum_{i=1}^{\mathcal{N}}{\sum_{j=1}^{\mathcal{N}}{a_{ij}\hat{c}_{ij}m_{ij}}}\left\| C_ie_{ji}^{w} \right\| ^2
		\nonumber\\
		&+(2\alpha+\frac{2}{d_M}\bar{\eta}\gamma _c)\bar{m}\tilde{\delta}^TC^T(\widehat{L}\otimes I)C\tilde{\delta}
		\nonumber\\
		&+\frac{1}{2}\rho \vartheta \lambda _{2}^{-1}\bar{m}\tilde{\delta}^TC^T(L\otimes I)C\tilde{\delta}
	+\frac{1}{4}\beta \bar{m}\tilde{w}^TC^T(L\otimes I)C\tilde{w}\label{e40}
	\end{align}
\end{lemma}

\emph{Proof}: At first,	taking the set-valued Lie derivative of $V_2$ in \eqref{e25th_b} along the time yields
\begin{align}\label{e35}
	\ell \dot{V}_2=&\frac{1}{d_M}\mathbb{F} \bigg\{\sum_{i=1}^{\mathcal{N}}{\sum_{j=1}^{\mathcal{N}}{a_{ij}\dot{m}_{ij}\left( \left\| e_{ji}^{\delta} \right\| ^2+\left\| e_{ji}^{w} \right\| ^2 + \| e_{ji}^{c} \|^2 \right)}}
	\nonumber\\
	&+\rho \sum_{i=1}^{\mathcal{N}}{\sum_{j=1}^{\mathcal{N}}{a_{ij}m_{ij} e_{ji}^{\delta T}C_{i}^{T}\mathbf{f}_i}}
	\nonumber\\
	&+\alpha \sum_{i=1}^{\mathcal{N}}{\sum_{j=1}^{\mathcal{N}}{\sum_{j=1}^{\mathcal{N}}{a_{ij}\hat{c}_{ij}m_{ij}{e_{ji}^{\delta}}^T}C_{i}^{T}\left( C_i\hat{\delta}_{ji}-C_j\hat{\delta}_{ij} \right)}}
	\nonumber\\
	&+\beta \sum_{i=1}^{\mathcal{N}}{\sum_{j=1}^{\mathcal{N}}{\sum_{j=1}^{\mathcal{N}}{a_{ij}m_{ij}{e_{ji}^{\delta}}^T}}}C_{i}^{T}\left( C_i\hat{w}_{ji}-C_j\hat{w}_{ij} \right) 
	\nonumber\\
	&-\alpha \sum_{i=1}^{\mathcal{N}}{\sum_{j=1}^{\mathcal{N}}{\sum_{j=1}^{\mathcal{N}}{a_{ij}\hat{c}_{ij}m_{ij}{e_{ji}^{w}}^T}C_{i}^{T}}\left( C_i\hat{\delta}_{ji}-C_j\hat{\delta}_{ij} \right)}\nonumber\\
	&+\sum_{i=1}^{\mathcal{N}}{\sum_{j=1}^{\mathcal{N}}{a_{ij}m_{ij}\eta_{ij}}}e_{ij}^{c}\| C_i\hat{\delta}_{ji}-C_j\hat{\delta}_{ij}\|^2 \bigg\}
\end{align}
Employing the conditions that $C_i\hat{\delta}_{ji}-C_j\hat{\delta}_{ij}=( C_ie_{ji}^{\delta}-C_je_{ij}^{\delta} ) +( C_i\tilde \delta _i-C_j \tilde \delta _j )$,
$C_i\hat{w}_{ji}-C_j\hat{w}_{ij}=( C_ie_{ji}^{w}-C_je_{ij}^{w}) + ( C_i\tilde w_i-C_j\tilde w_j )$, and using Young's inequality theorem, we have
$$\begin{aligned}
{e_{ji}^{\delta}}^T&C_{i}^{T}\left( C_i\hat{\delta}_{ji}-C_j\hat{\delta}_{ij} \right) 
\\
\leqslant& \frac{1}{2}\left\| C_ie_{ji}^{\delta} \right\| ^2+\frac{1}{2}\left\| C_je_{ij}^{\delta} \right\| ^2+\frac{1}{2}\left\| C_ie_{ji}^{\delta} \right\| ^2
\\
&+\frac{1}{2}\left\| C_ie_{ji}^{\delta} \right\| ^2+\frac{1}{2}\left\| C_i\tilde{\delta}_i-C_j\tilde{\delta}_j \right\| ^2,
\\
e_{ji}^{\delta T}&C_{i}^{T}\left( C_i\hat{w}_{ji}-C_j\hat{w}_{ij} \right) 
\\
\leqslant& \frac{1}{2}\left\| C_ie_{ji}^{\delta} \right\| ^2+\frac{1}{2}\left\| C_ie_{ji}^{w} \right\| ^2+\frac{1}{2}\left\| C_je_{ij}^{w} \right\| ^2
\\
&+2\left\| C_je_{ij}^{\delta} \right\| ^2+\frac{1}{8}\left\| C_iw_i-C_jw_j \right\| ^2,
\end{aligned}$$
$$\begin{aligned}
	-{e_{ji}^{w}}^T&C_{i}^{T}\left( C_i\hat{\delta}_{ji}-C_j\hat{\delta}_{ij} \right) 
	\\
	\leqslant& \frac{1}{2}\left\| C_ie_{ji}^{w} \right\| ^2+\frac{1}{2}\left\| C_je_{ij}^{\delta} \right\| ^2+\frac{1}{2}\left\| C_ie_{ji}^{\delta} \right\| ^2
	\\
	&+\frac{1}{2}\left\| C_ie_{ji}^{w} \right\| ^2+\frac{1}{2}\big\| C_i\tilde{\delta}_i-C_j\tilde{\delta}_j \big\| ^2,\\
		e_{ij}^{c}\|& C_i\hat{\delta}_{ji}-C_j\hat{\delta}_{ij}\|^2
	\\
	\leqslant& \gamma_{c}\left( \|C_ie_{ji}^{\delta}\|^2+\parallel C_je_{ij}^{\delta}\parallel ^2 \right) +\gamma_{c} \|C_i\tilde{\delta}_i-C_j\tilde{\delta}_j\|^2
\end{aligned}$$
In addition, the following condition holds 
$$\begin{aligned}
	&\frac{1}{d_M} \rho\sum_{i=1}^{\mathcal{N}}{\sum_{j=1}^{\mathcal{N}}{a_{ij}m_{ij}{e_{ji}^{\delta}}^TC_{i}^{T} \mathbf{f}_i}}
	\\
	\leqslant& \frac{1}{2d_M}\rho \sum_{i=1}^{\mathcal{N}}{\sum_{j=1}^{\mathcal{N}}{a_{ij}m_{ij}\left( \left\| C_ie_{ji}^{\delta} \right\| ^2+\vartheta \|C_i {\delta}_{i} \| ^2 \right)}}
	\\
	\leqslant& \frac{1}{2d_M}\rho  \sum_{i=1}^{\mathcal{N}}{\sum_{j=1}^{\mathcal{N}}{a_{ij}m_{ij}\left\| C_ie_{ji}^{\delta} \right\| ^2}}
	\\
	&+\frac{1}{2}\rho \vartheta \lambda _{2}^{-1}\bar m \tilde{\delta}^TC^T(L\otimes I)C\tilde{\delta}
\end{aligned}$$	

Consequently, one can derive
$$\begin{aligned}
\ell \dot{V}_2&\leqslant \frac{1}{d_M}\sum_{i=1}^{\mathcal{N}}{\sum_{j=1}^{\mathcal{N}}{a_{ij}\dot{m}_{ij}\lambda _{M}^{-2}\left( C \right) \left( \left\| C_ie_{ji}^{\delta} \right\| ^2+\left\| C_je_{ji}^{w} \right\| ^2 \right)}}\\
&+\left(3\alpha+\frac{5}{2}\beta + \frac{2\bar \eta \gamma_{c}}{d_M}+\frac{\rho }{2d_M}\right)\sum_{i=1}^{\mathcal{N}}{\sum_{j=1}^{\mathcal{N}}{a_{ij}\hat c_{ij}m_{ij}}}\left\| C_ie_{ji}^{\delta} \right\| ^2\\
&+(\alpha+\beta) \sum_{i=1}^{\mathcal{N}}{\sum_{j=1}^{\mathcal{N}}{a_{ij}\hat c_{ij}m_{ij}}}\left\| C_ie_{ji}^{w} \right\| ^2\\
&+\frac{1}{2}\rho \vartheta \lambda _{2}^{-1}\bar m \tilde{\delta}^TC^T(L\otimes I)C\tilde{\delta}
\\
&+\left(\alpha + \frac{1}{d_M} \bar \eta \gamma_{c}\right)\sum_{i=1}^{\mathcal{N}}{\sum_{j=1}^{\mathcal{N}}{a_{ij}\hat c_{ij}m_{ij}}} \| C_i\tilde{\delta}_i-C_j\tilde{\delta}_j \| ^2\\
&+\frac{1}{8}\beta \sum_{i=1}^{\mathcal{N}}{\sum_{j=1}^{\mathcal{N}}{a_{ij}m_{ij}}}\left\| C_i\tilde{w}_i-C_j\tilde{w}_j \right\| ^2
\end{aligned}$$	 
which indicates that the inequality \eqref{e40} holds. This completes the proof. $\hfill\blacksquare$

\subsection{Proof of Theorem 2 } \label{appendixC}

Define $\dot{\hat{\chi}}_{i,1}=A_{i,12}K_i\hat{\chi}_{i,1}-A_{i,11}\tilde{\chi}_{i,1}+A_{i,12}\chi_{i,2}$, and $\tilde{\chi}_{i,1}=\hat{\chi}_{i,1}-\chi_{i,1}$. That implies $\dot{\tilde{\chi}}_{i,1}=\varLambda _i\tilde{\chi}_{i,1}$. Since $K_i$ is of full column rank that ensures $\varLambda _i$ is Hurwitz, it makes $\lim _{t\rightarrow \infty}\tilde{\chi}_{i,1}=0$. Then, we design the Lyapunov candidate function $V_3=V_4+V_5$ with $V_4=\chi_{i,1}^{T}P_i\chi_{i,1}$, $V_5=\chi_{i,2}^{T}\chi_{i,2}$. Due to the conditions $\chi_{i,1}=x_{i,1}-x_{i,1}^{*}$, $\hat \chi_{i,1}=\hat x_{i,1}-x_{i,1}^{*}$, $\chi_{i,2}=-\bar{K}_i\left( x_i-\delta _i \right)$, we have
$$\begin{aligned}
	\dot{V}_4=&2\hat{\chi}_{i,1}^{T}P_i\left( \varLambda _i \hat{\chi}_{i,1}-A_{i,11}\tilde{\chi}_{i,1}+A_{i,12}\chi_{i,2} \right) 
	\\
	=&\hat{\chi}_{i,1}^{T}\left( \varLambda _i^TP_i+P_i\varLambda _i \right) \hat{\chi}_{i,1}+2\hat{\chi}_{i,1}^{T}P^{-1}A_{i,11}\tilde{\chi}_{i,1}\\
	&+2\hat{\chi}_{i,1}^{T}P^{-1}A_{i,12}\chi_{i,2}
	\\
	\leqslant& \hat{\chi}_{i,1}^{T}\left( \varLambda _i^T P_i+P_i\varLambda _i+2P_iP_i \right) \hat{\chi}_{i,1}
	\\
	&+\tilde{\chi}_{i,1}^{T}A_{i,12}^{T}A_{i,12}\tilde{\chi}_{i,1}+\chi_{i,2}^{T}A_{i,12}^{T}A_{i,12}\chi_{i,2}
	\\
	\leqslant&-\bar \mu \left\| \hat{\chi}_{i,1} \right\| ^2+\tilde{\chi}_{i,1}^{T}A_{i,12}^{T}A_{i,12}\tilde{\chi}_{i,1}+\chi_{i,2}^{T}A_{i,12}^{T}A_{i,12}\chi_{i,2}
\end{aligned} 
$$
Additionally, taking the derivative of $V_5$ obtains
$$\begin{aligned}
	\dot{V}_5=&\chi_{i,2}^{T} ( \dot{x}_{i,2}-\dot{\delta}_{i,2} ) -\chi_{i,2}^{T}K_i ( \dot{x}_{i,1}-\dot{\delta}_{i,1} ) 
	\\
	=&\chi_{i,2}^{T}[A_{i,21}x_{i,1}+A_{i,22}x_{i,2}+\bar{B}_iu_i-\varphi _{i,2}
	\\
	&-K_i\left( A_{i,11}x_{i,1}+A_{i,12}x_{i,2}-\varphi _{i,1} \right) ]
	\\
	=&\chi_{i,2}^{T}\left[ \bar{B}_iu_i-\bar{K}_i\left( A_ix_i-\varphi_i \right) \right]
	\\
	=&\chi_{i,2}^{T}F_i\bar{K}\left( x_i-\delta _i \right) 
	\\
	&+\chi_{i,2}^{T}\left[ -\left( \bar{K}_i\varphi _i-\bar{K}_iA_ix_i \right) -\bar{K}_i\left( A_ix_i-\varphi _i \right) \right] 
	\\
	=&-\chi_{i,2}^{T}F_i\chi_{i,2}
	\\
	=&-\bar \mu \chi_{i,2}^{T}\chi_{i,2}-\chi_{i,2}^{T}A_{i,12}^{T}A_{i,12}\chi_{i,2}
\end{aligned} 
$$
Then, it is not difficult to obtain that
$$\begin{aligned}
	\dot{V}_3\leqslant -\bar \mu \left\| \hat{\chi}_{i,1} \right\| ^2- \bar \mu \left\| \chi_{i,2} \right\| ^2+\tilde{\chi}_{i,1}^{T}A_{i,12}^{T}A_{i,12}\tilde{\chi}_{i,1}
\end{aligned}
$$
Since $\lim _{t\rightarrow \infty}\tilde{\chi}_{i,1}=0$, then invoking Barbalat's Lemma, we have that $\hat{\chi}_{i,1}$ and ${\chi}_{i,2}$ converge to zero exponentially. Given that $\lim _{t\rightarrow \infty}\hat{z}_{i,1}=z_{i,1}$, one can straightforwardly obtain that $\lim _{t\rightarrow \infty} \chi_{i,1}=0$. It indicates
$\lim _{t\rightarrow \infty}x_i=\delta_{i}$.
Therefore, by Theorem \ref{theoreom.1a}, one obtains
\begin{align}\label{e8}
	\lim _{t\rightarrow \infty}C_ix_{i}=C_i\delta_i^{*}
\end{align}
From Lemma \ref{lamma6}, this implies that  $\lim_{t\rightarrow \infty} y_i={y}_i^*=y_{j}^{*}$, $\forall i,j \in \mathcal{I_N}$, and ${y}_i^*$ is the optimal solution of Problem \ref{pro1}.

		\small{
			\bibliographystyle{IEEEtran}
			\bibliography{references}
		}
		%\section*{References}
		
		%\begin{IEEEbiography}[{\includegraphics[width=1in,height=1.25in,clip,keepaspectratio]{a1.png}}]{First A. Author} (Fellow, IEEE) and all authors may include 
		%biographies. Biographies are
		%often not included in conference-related papers.
		%This author is an IEEE Fellow. The first paragraph
		%may contain a place and/or date of birth (list
		%place, then date). Next, the author’s educational
		%background is listed. The degrees should be listed
		%with type of degree in what field, which institution,
		%city, state, and country, and year the degree was
		%earned. The author’s major field of study should
		%be lower-cased.
		
		%The second paragraph uses the pronoun of the person (he or she) and
		%not the author’s last name. It lists military and work experience, including
		%summer and fellowship jobs. Job titles are capitalized. The current job must
		%have a location; previous positions may be listed without one. Information
		%concerning previous publications may be included. Try not to list more than
		%three books or published articles. The format for listing publishers of a book
		%within the biography is: title of book (publisher name, year) similar to a
		%reference. Current and previous research interests end the paragraph.
		
		%\end{IEEEbiography}
		
		%\begin{IEEEbiographynophoto}{Second B. Author,} photograph and biography not available at the
		%time of publication.
		%\end{IEEEbiographynophoto}
		
		%\begin{IEEEbiographynophoto}{Third C. Author Jr.} (Member, IEEE), photograph and biography not available at the
		%time of publication.
		%\end{IEEEbiographynophoto}
		
	\end{document}